\documentclass[letter]{article}

\usepackage{graphicx,amsmath,url,pdfpages}
\usepackage[margin=1in]{geometry}

\begin{document}

\title{\bf BPPart and BPMax: RNA-RNA Interaction Partition Function and Structure
  Prediction for the Base Pair Counting Model}

\author{%
Ali Ebrahimpour-Boroojeny,\,
Sanjay Rajopadhye, and\,
 Hamidreza Chitsaz
\footnote{To whom correspondence should be addressed.
Email: chitsaz@chitsazlab.org}
\\\ \\
Department of Computer Science, Colorado State University}


\date{}

\maketitle

\begin{abstract}
  A few elite classes of RNA-RNA interaction (RRI), with complex roles in cellular functions such as miRNA-target and lncRNAs in human health, have already been studied. Accordingly, RRI bioinformatics tools tailored for those elite classes have been proposed in the last decade.   Interestingly, there are somewhat unnoticed mRNA-mRNA interactions in the literature with potentially drastic biological roles.  Hence, there is a need for high-throughput \emph{generic} RRI bioinformatics tools.

  We revisit our RRI partition function
  algorithm, \texttt{piRNA}, which happens to be the most comprehensive and
 computationally-intensive thermodynamic model for RRI. We propose simpler models that are shown to retain the vast majority of the
  thermodynamic information that \texttt{piRNA} captures. 

  We simplify the energy model and instead consider only weighted base pair counting to obtain \texttt{BPPart} for Base-pair Partition function and \texttt{BPMax} for Base-pair Maximization which are $225\times$ and $1350\times$ faster than \texttt{piRNA}, with a correlation of 0.855 and 0.836 with \texttt{piRNA} at
  $37^{\circ}C$ on 50,500 experimentally characterized RRIs.  This correlation
  increases to 0.920 and 0.904, respectively, at $-180^{\circ}C$.
  
  Finally, we apply our algorithm \texttt{BPPart} to discover two disease-related RNAs, SNORD3D and TRAF3, and hypothesize their potential roles in Parkinson's disease and Cerebral Autosomal Dominant Arteriopathy with Subcortical Infarcts and Leukoencephalopathy (CADASIL).

\end{abstract}

\section{Introduction}


Since mid 1990s with the advent of RNA interference discovery, RNA-RNA
interaction (RRI) has moved to the spotlight in modern, post-genome biology.
RRI is ubiquitous and has increasingly complex roles in cellular functions.
In human health studies, miRNA-target and lncRNAs are among an elite class of
RRIs that have been extensively studied and shown to play significant roles in
various diseases including cancer.  Bacterial ncRNA-target and RNA
interference are other classes of RRIs that have received significant
attention.  However, new evidence suggests that other classes of RRI, such as
mRNA-mRNA interactions, are biologically important.

The RISE database~\cite{Gong17} reports a number of biologically significant
instances of mRNA-mRNA interactions.  These representative mRNA-mRNA
interactions suggest that general RRIs, including mRNA-mRNA interactions, play
major roles in human biology.  Hence, there is a need for high-throughput
\emph{generic} RNA-RNA interaction bioinformatic tools for all types of RNAs.

\begin{figure*}[t]
\begin{center}
\includegraphics[angle=-90,width=0.40\textwidth]{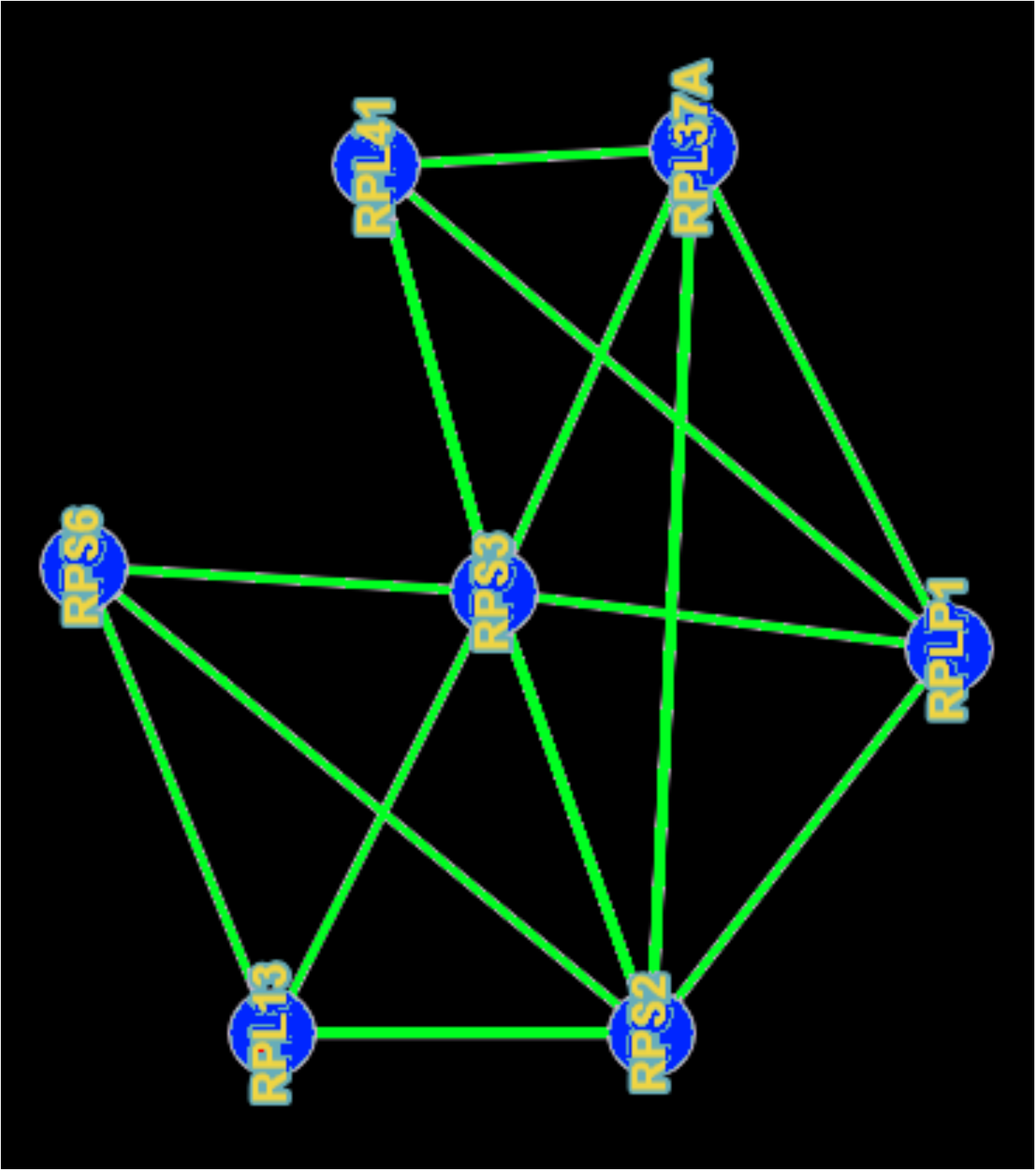}
\end{center}
\caption{A substructure of the genes in the ribosome pathway.  Each node
    represents a gene and each edge represents an experimentally observed
    interaction between the corresponding genes.}
\label{fig:graph}
\end{figure*}

As an example of this necessity for all types of RNAs, we found 3 cliques of
size 4 of interacting protein-coding RNAs in ribosome which conform to what we
generally expect from the structure of the ribosome.  These cliques are highly
entangled together to form an interaction graph as Figure~\ref{fig:graph}.
RPS3 which seems to be one of the genes with the highest number of connections
interacts with at least 14 other genes in ribosome pathway.  Another
interesting clique of size 4 that we could find consists of 4 genes in the
pathway of regulation of actin cytoskeleton, ACTB, ACTG1, PFN1, and TMSB4X.
These genes are involved in vital tasks of proliferation, migration, mobility,
and differentiation of the cell.  Being able to capture all the interactions
that RNAs might have will help us better understand the post-transcriptional regulation
of the genes.


In this paper, we revisit our RNA-RNA interaction partition function
algorithm, \texttt{piRNA}, which happens to be the most comprehensive, albeit
the most computationally intensive, thermodynamic model for RNA-RNA
interaction \cite{Chitsaz09}. \texttt{piRNA} is a dynamic programming
algorithm that computes the partition function, base-pairing probabilities,
and structure for the comprehensive Turner energy model in $O(n^4m^2+n^2m^4)$
time and $O(n^4+m^4)$ space.  Due to intricacies of the energy model,
including various (kissing) loops such as hairpin loop, bulge/internal loop, and
multibranch loop, \texttt{piRNA} involves 96 different dynamic programming
tables and needs multiple table look-ups for computing their values. An implementation of \texttt{piRNA} is currently available at 
\url{http://chitsazlab.org/software/pirna}.

In this paper, we introduce a strategic retreat from the slower comprehensive
models such as \texttt{piRNA} by simplifying the energy model and instead
considering only simple weighted base pair counting.  We develop the
\texttt{BPPart} algorithm for Base-pair Partition function and \texttt{BPMax}
for Base-pair Maximization, both of which are faster by a significant, albeit
constant factor.
By the explosion of experimental
data which makes us able to use machine learning methods, such as deep
learning, for detection of RNA subsequences that interact, this retreat is
necessary if one is willing to build physics-guided models by using the
features that are derived by an energy model.  \texttt{BPPart} involves nine
4-dimensional dynamic programming tables, and \texttt{BPMax} involves only one
4-dimensional table.  Both \texttt{BPPart} and \texttt{BPMax} compared with
\texttt{piRNA} are simpler dynamic programming algorithms which are more than
$225\times$ and $1300\times$ faster, respectively, on the 50,500~RRI samples
we used for our experiments.  The reason for this noticeable speed-up is
reducing the number of tables and the number of table look-ups for computing
the new values and also the fact that the 96 large tables of \texttt{piRNA}
renders \texttt{piRNA} memory- rather than compute-bound in practice.
Moreover, the significantly reduced memory footprint of \texttt{BPPart} and
\texttt{BPMax} makes them feasible targets for optimization on different
hardware platforms like GPU based accelerators, an avenue we plan to explore
in the future.

The key question concerns the accuracy we lose by simplifying the scoring
model from the comprehensive Turner model to simply weighted base-pair
counting.  We answer this by computing both the Pearson and Spearman's rank
correlations at different temperatures between the results of \texttt{BPPart},
\texttt{BPMax}, and \texttt{piRNA} on 50,500 experimentally characterized RRIs
in the RISE database~\cite{Gong17}.  We find that the Pearson correlations
between \texttt{BPPart} and \texttt{piRNA} is 0.920 and \texttt{BPMax} and
\texttt{piRNA} is 0.904 at $-180^{\circ}C$ after optimizing the weights for
base pairs.  The effect of entropy, for which the simple base pair counting
model does not account, increases with temperature.  Completely in conformance
with this theoretical expectation, we find that the Pearson correlations
between \texttt{BPPart} and \texttt{piRNA} is 0.855 and \texttt{BPMax} and
\texttt{piRNA} is 0.836 at $37^{\circ}C$.
We conclude that \texttt{BPPart} and \texttt{BPMax} capture a significant
portion of the thermodynamic information.  They can possibly be complemented
with machine learning techniques in the future for more accurate predictions.

\subsection{Related work}

During the last few decades, several computational methods emerged to study
the secondary structure of single and interacting nucleic acid strands.  Most
use a thermodynamic model such as the well-known Nearest Neighbor
Thermodynamic model \cite{MatTur99, CaoChe06, DirPie03, Chitsaz09, Nus78,
  WatSmi78, Zuker81, RivEdd99, Mcc90, wenzel2012risearch}.  Some previous
attempts to analyze the thermodynamics of multiple interacting strands
concatenate input sequences in some order and consider them as a single strand
\cite{AndZha05,BerTaf06,DirBoiSchWinPie07}.  Alternatively, several methods
avoid internal base-pairing in either strand and compute the minimum free
energy secondary structure for their hybridization under this constraint
\cite{RehSte04,DimZuk04,MarZuk08}.  The most comprehensive solution is
computing the joint structure between two interacting strands under energy
models with a growing complexity
\cite{Pervouchine04,AlkKarNadSahZha06,lorenz2011viennarna,dichiacchio2015accessfold,Kato09,Chitsaz09,Huang09}.

Other methods predict the secondary structure of individual RNA independently,
and predict the (most likely) hybridization between the unpaired regions of
the two interacting molecules as a multistep process: 1) unfolding of the two
molecules to expose bases needed for hybridization, 2) the hybridization at
the binding site, and 3) restructuring of the complex to a new minimum free
energy conformation \cite{Muckstein06,Walton02,BusRicBac08,Chitsaz09b}.  The
success of such methods, including our biRNA algorithm \cite{Chitsaz09b},
suggests that the thermodynamic information vested in subsequences and pairs
of subsequences of the input RNAs can provide valuable information for
predicting features of the entire interaction.

  
In addition to general RNA-RNA interaction tools, many tools have been
developed to predict the secondary structure of interacting RNAs for a
specific type of interest which has been shown to be more effective in some
cases due to the utilization of certain properties belonging to that type.  As
mentioned earlier, miRNA-target prediction is one such class of high interest
for which such specialized tools have been created to incorporate various
properties specific to miRNAs; some of these tools use the seed region of a
miRNA which is highly conserved
\cite{krek2005combinatorial,kertesz2007role,kruger2006rnahybrid,zhang2005miru},
some consider the free energy to compute accessibility to the binding site in
$3'$~UTR \cite{hofacker1994fast,lorenz2011viennarna,kertesz2007role}, some
utilize the conservation level which is derived using the phylogenetic
distance \cite{nam2014global, betel2010comprehensive, reczko2012functional,
  gaidatzis2007inference, krek2005combinatorial, kertesz2007role}, and some
others consider other target sites as well, such as the $5'$ UTR, Open Reading
Frames (ORF), and the coding sequence (CDS) for mRNAs \cite{
  riffo2016tools,miranda2006pattern, hsu2011mirtar, xu2014identifying}.

There are also several other tools developed for other specific types of RNA;
IntaRNA \cite{busch2008intarna, mann2017intarna} is one such tool that
although is used for RNA-RNA interaction in general, it is primarily designed
for predicting target sites of non-coding RNAs (ncRNAs) on mRNAs.  There are
many other examples, such as PLEXY \cite{kehr2010plexy} which is a tool
designed for C/D snoRNAs, RNAsnoop \cite{tafer2010rnasnoop} that is designed
for H/ACA snoRNAs, TargetRNA \cite{tjaden2008targetrna} which is a tool aimed
at predicting interaction of bacterial sRNAs \cite{umu2017comprehensive}.

\section{Methods}

\newcommand{\RR}{{\bf R}}
\newcommand{\RS}{{\bf S}}

Here we describe how our algorithm, \texttt{BPPart}, utilizes a dynamic
programming approach to compute the partition function for RNA-RNA interaction
when entropy is ignored and only a weighted score for pairing different
nucleotides is considered.  This algorithm is guaranteed to be mutually
exclusive on the set of structures, i.e., it counts each structure exactly
once.  For \texttt{BPMax} which maximizes the (weighted scores) of base-pairs,
such mutual exclusion is not necessary because the $\max$ operator is
idempotent (counting the same structure multiple times does not affect the
value of the objective function) and we give a $10\times$ simpler recursion.
Our codes are freely available under open source license.\footnote{See
  \texttt{https://github.com/Ali-E/bipart}}

\subsection*{Preliminaries}

In this paper, we mostly follow the notations and definitions used to develop
our \texttt{piRNA} algorithm~\cite{Chitsaz09}.  We denote the two nucleic acid
strands by $\RR$ and $\RS$.  Strand $\RR$ is indexed from $1$ to $L_R$, and
$\RS$ is indexed from $1$ to $L_S$ both in $5'$ to $3'$ direction.  Note that
the two strands interact in opposite directions, e.g. $\RR$ in
$5' \rightarrow 3'$ with $\RS$ in $3' \leftarrow 5'$ direction; however, we
consider the reverse of $\RS$ in our figures for clearer illustration of the
configurations.  Each nucleotide is paired with at most one nucleotide in the
same or the other strand. The subsequence from the $i^{th}$ nucleotide to the
$j^{th}$ nucleotide, inclusive, in either strand is denoted by $[i, j]$.

An intramolecular base pair between the nucleotides $i$ and $j$ (by
convention, $i<j$) in a strand is called an {\it arc} and denoted by a bullet
$i \bullet j$.  We represent the score of such arc by $\mathrm{score}(i,j)$. Essentially, $\mathrm{score}(i,j)$ is $c_1$ if $i \bullet j$ is GU or UG, is $c_2$ if $i \bullet j$ is AU or UA, and is $c_3$ if $i \bullet j$ is CG or GC. 
An intermolecular base pair between the nucleotides $k_1$ and $k_2$, where
$k_1 \in R, k_2 \in S$, is called a \emph{bond}, denoted by a circle
$k_1 \circ k_2$.  We represent the score of such a bond by
$\mathrm{iscore}(k_1,k_2)$.  
Essentially, $\mathrm{iscore}(k_1,k_2)$ is $c'_1$ if $k_1 \circ k_2$ is GU or UG, is $c'_2$ if $k_1 \circ k_2$ is AU or UA, and is $c'_3$ if $k_1 \circ k_2$ is CG or GC.

\begin{figure*}[t]
\begin{center}
\begin{tabular}{cc}
\includegraphics[width=0.94\textwidth]{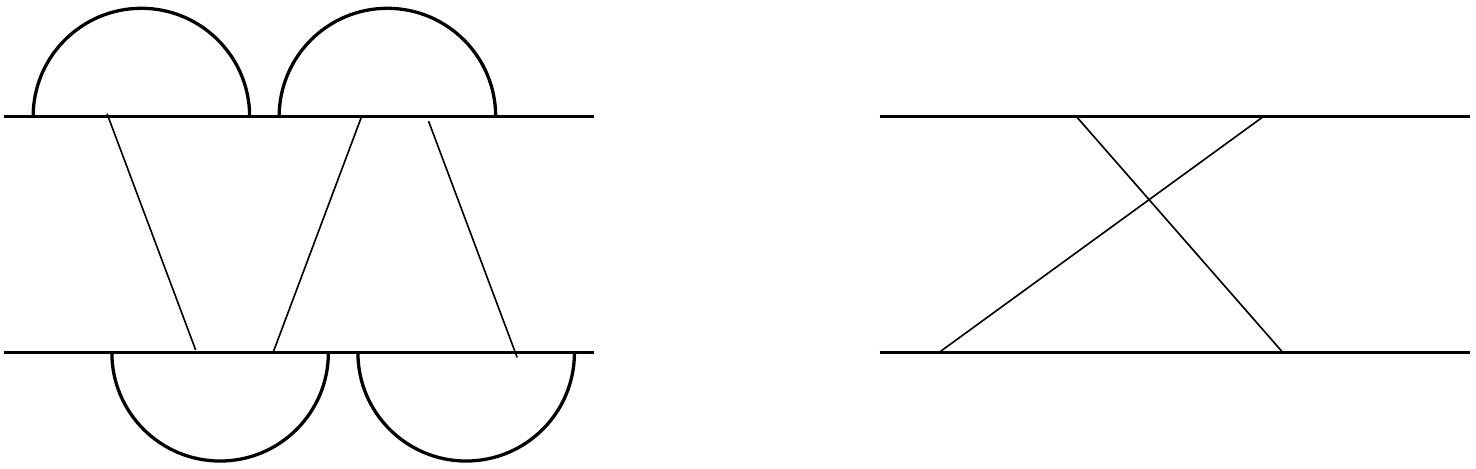}
\end{tabular}
\end{center}
\caption{An illustration of a zigzag (left) and a crossing bond (right), which are excluded in our algorithm.}
\label{fig:zigzag}
\end{figure*}

An arc $i \bullet j$ in R \emph{covers} a bond
$k_1 \circ k_2$ if $i_1 < k_1 < j_1$.  We call $i \bullet j$ an
\emph{interaction arc} in R if there is a bond $k_1 \circ k_2$ covered by
$i \bullet j$.  The \emph{scope} of an interaction arc is the interval
$[i+1, j-1]$.  We call a base on either strand an \emph{event} if it is either
the end point of a bond or that of an interaction arc. In our explanation we may use
arc and bond as verbs. Two bonds $i_1 \circ i_2$ and $j_1 \circ j_2$ are called \emph{crossing bonds} (right case of Figure~\ref{fig:zigzag})
if $i_1 < j_1$ and $i_2 > j_2$, or vice versa.  An interaction arc
$i_1 \bullet j_1$ in a strand \emph{subsumes} a subsequence $[i_2, j_2]$ in
the other strand if none of the bases in $[i_2, j_2]$ has a bond with a base
outside this arc.  Mathematically, for all bonds $k_1 \circ k_2$ where
$i_2<k_2<j_2$, $k_1$ lies within the scope of $i_1\bullet j_1$.  Two
interaction arcs are \emph{equivalent} if they subsume one another.  Two
interaction arcs $i_1 \bullet j_1$ and $i_2 \bullet j_2$ are part of a
\emph{zigzag}, if neither $i_1 \bullet j_1$ subsumes $[i_2, j_2]$ nor
$i_2 \bullet j_2$ subsumes $[i_1, j_1]$ (left case of Figure~\ref{fig:zigzag}).

In this work, we assume there are no pseudoknots in individual secondary
structures of $\RR$ and $\RS$, and also there are no crossing bonds and
no zigzags between $\RR$ and $\RS$.  These constraints allow a polynomial
algorithm---the general case of considering all possible structures is NP-hard
\cite{AlkKarNadSahZha06}. We denote the ensemble of unpseudoknotted structures of $\RR$ and $\RS$ by
$\mathcal{S}(\RR)$ and $\mathcal{S}(\RS)$ respectively.  The ensemble of
unpseudoknotted, crossing-free, and zigzag-free joint interaction structures
is denoted by $\mathcal{S}^I(\RR,\RS)$.

For a given structure $s$ in either $\mathcal{S}(\RR) $ or $\mathcal{S}(\RS)$,
let $\mathrm{AU}(s)$ denote the number of A-U base pairs in $s$.  Similarly,
$\mathrm{CG}(s)$ and $\mathrm{GU}(s)$ denote the number of C-G and G-U base
pairs in $s$, respectively.  We define \emph{bpcount} as a weighted sum, for
some constants, $c_1,\ldots, c_3$
\begin{equation}
  \mathrm{bpcount}(s) = c_1 \mathrm{GU}(s) + c_2 \mathrm{AU}(s) + c_3
  \mathrm{CG}(s).
\end{equation}

For a given joint interaction structure $s \in \mathcal{S}^I(\RR,\RS)$, let
$\mathrm{AU}(s)$, $\mathrm{CG}(s)$, and $\mathrm{GU}(s)$ denote the respective
number of intramolecular base pairs in $s$, and let $\mathrm{AU}^I(s)$,
$\mathrm{CG}^I(s)$, and $\mathrm{GU}^I(s)$ denote the number of corresponding
intermolecular base pairs in $s$.  We define for
some constants, $c'_1,\ldots, c'_3$, for any joint interaction
structure $s$, 
\begin{equation}
\mathrm{bpcount}^I(s) = c'_1 \mathrm{GU}^I(s) + c'_2 \mathrm{AU}^I(s) + c'_3
\mathrm{CG}^I(s),
\end{equation}
and
\begin{equation}
\mathrm{bpcount}(s) = c_1 \mathrm{GU}(s) + c_2 \mathrm{AU}(s) + c_3 \mathrm{CG}(s) + \mathrm{bpcount}^I(s).
\end{equation} 

\subsection*{Rivas-Eddy Diagrams}
\enlargethispage{-60.1pt}
For the sake of completeness, we describe the ``Rivas-Eddy diagram'' notation that
we adopt in this paper.  The main elements are:

\begin{enumerate} \itemsep 0mm
\item A solid horizontal straight line represents a sequence; we have two
  sequences drawn as two parallel horizontal lines.
\item A solid curved line between two points in the same sequence is an arc;
  all arcs are either above the upper sequence, or below the lower one.
\item A dotted curved line with a cross in the middle, between two points in
  the same sequence means that those two points \emph{do not} form an arc.
\item A dashed curved line between two points in the same sequence denotes
  either 2 or 3.
\item A solid line between two points in different sequences is a bond.
\item Similarly, a dotted line with a cross in the middle, between two points
  in different sequences means that those two points \emph{do not} form a
  bond.
\item A dashed line between two points in different sequences denotes either 5
  or 6.
\item A region is the space under an arc, or between bonds.  When there are no
  additional choices of bonds/arcs in a given region, we fill it with a color
  (cyan); no arc or bond crosses a filled region.
\item A point in a sequence may be labeled with an index, and in general, the
  set of such indices are free variables used in the recursions; the index of
  unlabeled points before (after) labeled points is assumed to be the
  predecessor (successor) of the label.
\item A diagram may be labeled with the name(s) of the constituent
  substructures (which are eventually implemented as dynamic programming
  tables/variables).
\item A vanishing arc (i.e., one that starts at some index, and does not
  explicitly specify an end point) represents a structure whose start point is
  as specified, and the end point is to be determined.
\end{enumerate}

The Rivas-Eddy diagram to compute a certain function is written like a formal
(context free) grammar.  The left hand side is labeled with the name of a
table (structure), and the right hand side has a number of alternate
substructures separated by vertical bars.  Often, some of the boundary cases
(e.g., singleton or empty subsequences) are omitted for brevity.

\subsection*{Problem Definition}

In this paper, we solve two problems:
\begin{enumerate}
\item{\textbf{Base Pair Counting Partition Function:}} we give a dynamic
  programming algorithm \texttt{BPPart} to compute the partition function

\begin{equation}
\mathrm{Q}(\RR,\RS) = \sum_{s\in {\mathrm{S}^I(\RR,\RS)}} e^{\mathrm{bpcount}(s)},
\end{equation}

\item{\textbf{Base Pair Maximization:}} we give a dynamic programming
  algorithm \texttt{BPMax} to find the structure that has the maximum weighted
  base pair count, i.e.
\begin{equation}
  \mathrm{BPMax}(\RR,\RS) = \max_{s\in {\mathrm{S}^I(\RR,\RS)}}\;\;
  \mathrm{bpcount}(s).
\end{equation}
This problem was previously studied by Pervouchine \cite{Pervouchine04} in an
algorithm called IRIS.  However, there is no publicly available correct
implementation of IRIS.  Moreover, we also define an additional interaction
score to capture the structure with the highest intermolecular score, amongst
those that maximize the total score.  Mathematically,
\begin{equation}
  \mathrm{IS}(\RR,\RS) = \max_{\{s\ \mid\ \mathrm{bpcount}(s) =
    \mathrm{BPMax}(\RR,\RS)\}}
  \mathrm{bpcount}^I(s).
\end{equation}
We compute $\mathrm{IS}(\RR,\RS)$ by backtracing all possible
total-score-optimal structures, and selecting the one that has the maximum
intermolecular score.
\end{enumerate}

\subsection*{BPMax Algorithm}

We first explain the \texttt{BPMax} algorithm.  It is simpler than
\texttt{BPPart}, and allows us to describe the notation and conventions. When explaining some of the equations, helper functions, called $H,L,M,N$, are used to ease the reading of the paper. To differentiate these helper functions, superscripts are used.


For a single strand of nucleotides, we define $\mathrm{S}_{i,j}$ as the
maximum weighted sum of base pair scores on all possible foldings of
subsequence $[i,j]$.  We need to make such a table, for each of the $\RR$ and
$\RS$ strands, and we use superscripts $(1)$ and $(2)$, respectively, to
distinguish between them.  We also define $F_{i_1,j_1,i_2,j_2}$ as the maximum
weighted sum of base pair scores (considering both intra- and inter-pairings)
of subsequences $[i_1,j_1]$ from $\RR$ and $[i_2,j_2]$ from $\RS$.

The computation of $\mathrm{S}_{i,j}$ is based on the well known single RNA
folding algorithm~\cite{NusJac80}.  For short sequences (i.e., those whose
length is strictly less than 5) the score is $0$, otherwise, we use the
recursion in the second case of Equation~(\ref{equ_S}) shown below.  It
considers the case where we have an arc $i \bullet j$ and recurs on
$[i+1,j-1]$, and also other cases in which the $i^{th}$ and $j^{th}$ bases are
not paired and the $[i,j]$ is split into two smaller subsequences:

\begin{equation}
\label{equ_S}
\mathrm{S}_{i,j} = 
\left\{
  \begin{array}{ll}
    0 & j-i<4 \\
    \max \left(\mathrm{S}_{i+1,j-1} + \mathrm{score}(i,j),
    \;\displaystyle\max_{k=i}^{j-1} \mathrm{S}_{i,k} +
    \mathrm{S}_{k+1,j}\right)
      & \mathrm{otherwise.}
  \end{array} \right.
\end{equation}

\begin{figure*}[t]
\begin{center}
\begin{tabular}{cc}
\includegraphics[width=0.94\textwidth]{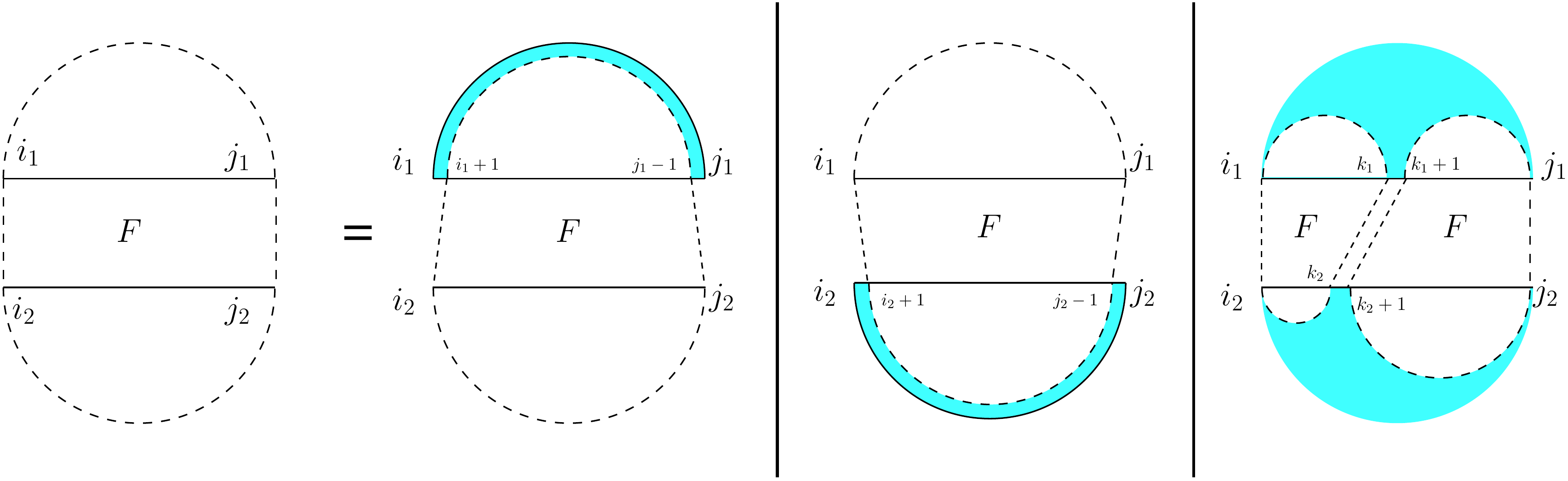}
\end{tabular}
\end{center}
\caption{The four cases defining table $F$.  Note that in the \texttt{BPMax}
  algorithm, the cases do not have to be mutually exclusive since we are
  working with the $\max$ operator, which is idempotent.}
\label{fig:F}
\end{figure*}

We define the recurrences for $\mathrm{F}_{i_1,j_1,i_2,j_2}$ similarly.  When
either sequence is empty, the value is simply the $\mathrm{S}$ of the other
sequence, and for two singleton sequences, it is the score of the single bond
possible.  Otherwise we have three cases: (i) $i_1$ arcs $j_1$
($i_1 \bullet j_1$) in which case the residual structure is given by a
recursion on $\mathrm{F}_{i_1+1,j_1-1,i_2,j_2}$, (ii) the symmetric case of
$i_2 \bullet j_2$ and $\mathrm{F}_{i_1,j_1,i_2+1,j_2-1}$, or (iii) none of
these arcs, and two recursive cases of $\mathrm{F}_{i_1,k_1,i_2,k_2}$ and
$\mathrm{F}_{k_1+1,j_1,k_2+1,j_2}$.  They are illustrated in the Rivas-Eddy diagram
of Figure~\ref{fig:F}, which lead to 

\begin{equation}
\label{equ_F}
\mathrm{F}_{i_1,j_1,i_2,j_2}= 
\left\{
\begin{array}{ll}
  -\infty & j_1 < i_1 \mbox{ and } j_2 < i_2 \\ \\
  \mathrm{S}^{(1)}_{i_1,j_1} & i_1 \leq j_1 \mbox{ and } j_2 < i_2  \\ \\
  \mathrm{S}^{(2)}_{i_2,j_2} & j_1 < i_1  \mbox{ and } i_2 \leq j_2\\ \\
  \mathrm{iscore}(i_1,i_2) & i_1 = j_1 \mbox{ and } i_2 = j_2  \\ \\
  \max \left[\; \mathrm{F}_{i_1+1,j_1-1,i_2,j_2} + \mathrm{score}(i_1,j_1), \right. \vspace*{1mm} \\ 
  \;\;\;\;\;\;\;\;\; \mathrm{F}_{i_1,j_1,i_2+1,j_2-1} + \mathrm{score}(i_2,j_2), \vspace*{1mm} \\ \;\;\;\;\;\;\;\;\; \left. H_{i_1,j_1,i_2,j_2} \; \right] & \mbox{otherwise,}\\ 
\end{array} \right.
\end{equation}

\begin{equation}
H_{i_1,j_1,i_2,j_2} = \displaystyle  \max_{k_1=i_1-1}^{j_1} \displaystyle
           \max_{k_2=i_2-1}^{j_2} (\mathrm{F}_{i_1,k_1,i_2,k_2} +
           \mathrm{F}_{k_1+1,j_1,k_2+1,j_2}).
\end{equation}
Note that $H$ is equivalent to
\begin{equation}
H_{i_1,j_1,i_2,j_2} = 
  \max\left(\begin{array}{l}
           \displaystyle \mathrm{S^{(1)}}(i_1,j_1) + \mathrm{S^{(2)}}(i_2,j_2),\\
           \displaystyle  \max_{k_1=i_1}^{j_1-1} \displaystyle
           \max_{k_2=i_2}^{j_2-1} \mathrm{F}_{i_1,k_1,i_2,k_2} +
           \mathrm{F}_{k_1+1,j_1,k_2+1,j_2}, \\ 
           \displaystyle \max_{k_2=i_2}^{j_2-1}
           \mathrm{S^{(2)}}(i_2,k_2) + \mathrm{F}_{i_1,j_1,k_2+1,j_2}, \\
           \displaystyle \max_{k_2=i_2}^{j_2-1}
           \mathrm{F}_{i_1,j_1,i_2,k_2} + \mathrm{S^{(2)}}(k_2+1,j_2), \\
           \displaystyle \max_{k_1=i_1}^{j_1-1}
           \mathrm{S^{(1)}}(i_1,k_1) + \mathrm{F}_{k_1+1,j_1,i_2,j_2}, \\
           \displaystyle \max_{k_1=i_1}^{j_1-1}
           \mathrm{F}_{i_1,k_1,i_2,j_2} + \mathrm{S^{(1)}}(k_1+1,j_1) \\
  \end{array}\right).
\label{helper_1}
\end{equation}

In Equation~(\ref{equ_F}), we compute $\mathrm{S}$ tables
separately for each strand, according to Equation (\ref{equ_S}) with the corresponding sequence as the input, and we distinguish them by superscripts
$^{(1)}$ and $^{(2)}$ above.  We use the same superscript convention throughout this paper.

\subsection*{BPPart Algorithm}

It is well known that the partition function can be computed by developing
similar recursions, with two simple modifications.  The first is that
algebraically, we operate with the field of reals rather than the
max-plus semi-ring.  Here, the additive identity is 0, rather than
\texttt{INT\_MIN} and the multiplicative identity is 1, rather than 0.  The
second, as already mentioned earlier, is that because addition is not
idempotent, we must carefully ensure that we enumerate substructures in a
mutually exclusive manner.

First, we start with the recursions for computing the partition function on a
single strand which is going to occur in many cases of the double-stranded
version.  Let $\mathrm{Q}_{i,j}$ represent the partition function of the
subsequence $[i, j]$.  As shown in Figure~\ref{fig:Q}, there are two mutually
exclusive cases: either (the right case) there is no arc, or (the left case)
there is a unique leftmost arc (the cyan fill ensures this) which starts at
$k$, and a substructure on $[k,j]$ with an arc starting at $k$, for which we
introduce a new table $\mathrm{Qz}$.

To define $\mathrm{Qz}_{i,j}$, let $i\bullet k$ (read as ``let $i$ arc $k$'')
for some index $k$.  This results in two $\mathrm{Q}$ substructures, one on
$[i+1,k-1]$, and the other on $[k+1,j]$.
Therefore, the value of $\mathrm{Qz}_{i,j}$ can be computed using
Equation~(\ref{equ_Qz}) which accounts for the assumption that no pairing is
allowed between two bases that are less than $3$ bases apart:

\begin{figure*}[t]
\begin{center}
\begin{tabular}{cc}
\includegraphics[width=0.75\textwidth]{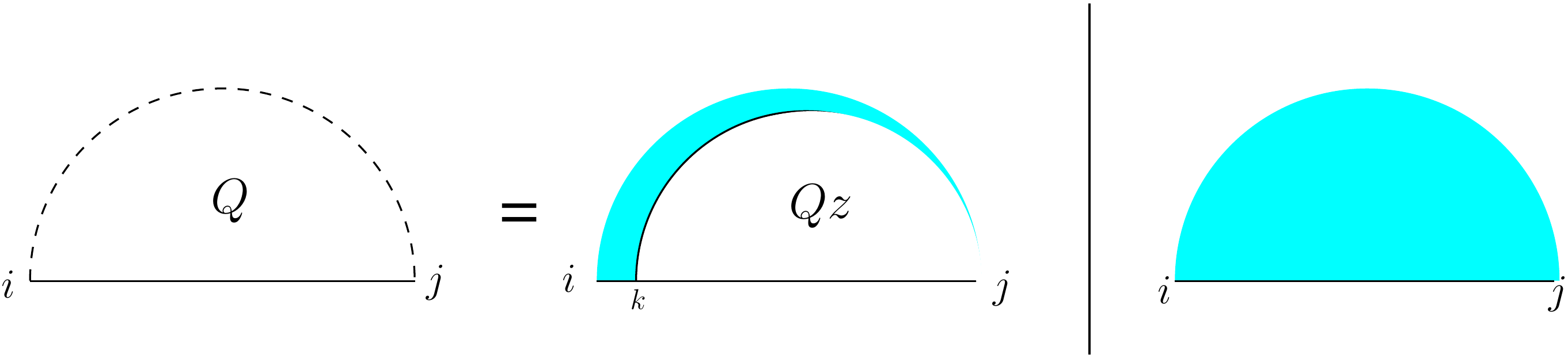}
\end{tabular}
\end{center}
\caption{For computing $\mathrm{Q}$, notice that either there is no pairing or
  there is at least one arc which starts at some index $k$ and results in a
  case of $\mathrm{Qz}$.}
\label{fig:Q}
\end{figure*}

\begin{figure*}[t]
\begin{center}
\begin{tabular}{cc}
\includegraphics[width=0.60\textwidth]{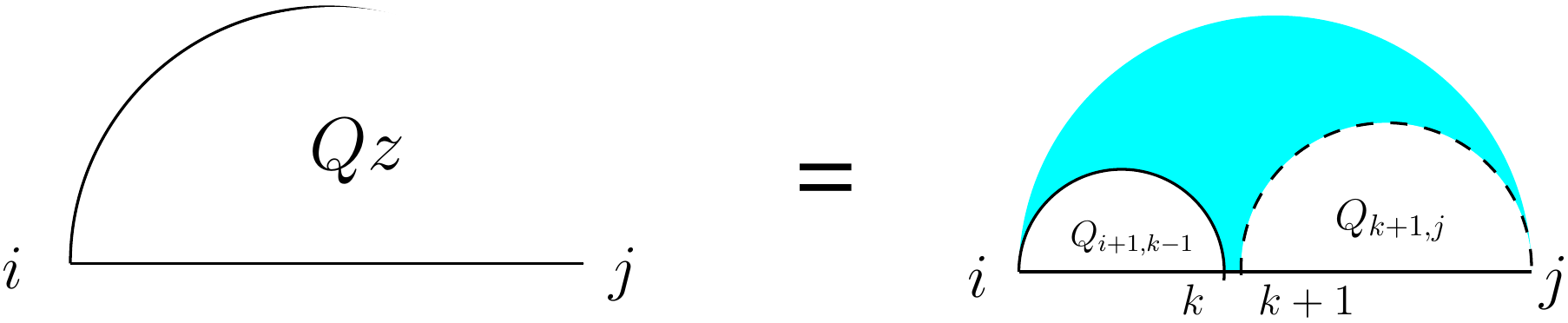}
\end{tabular}
\end{center}
\caption{Computing $\mathrm{Qz}$ can be achieved by considering the base $k$
  that is paired with $i$ and the two $\mathrm{Q}$ substructures it forms, one
  between $i$ and $k$ and one after $k$.}
\label{fig:Qz}
\end{figure*}

\begin{figure*}[t]
\begin{center}
\begin{tabular}{cc}
\includegraphics[width=0.9\textwidth]{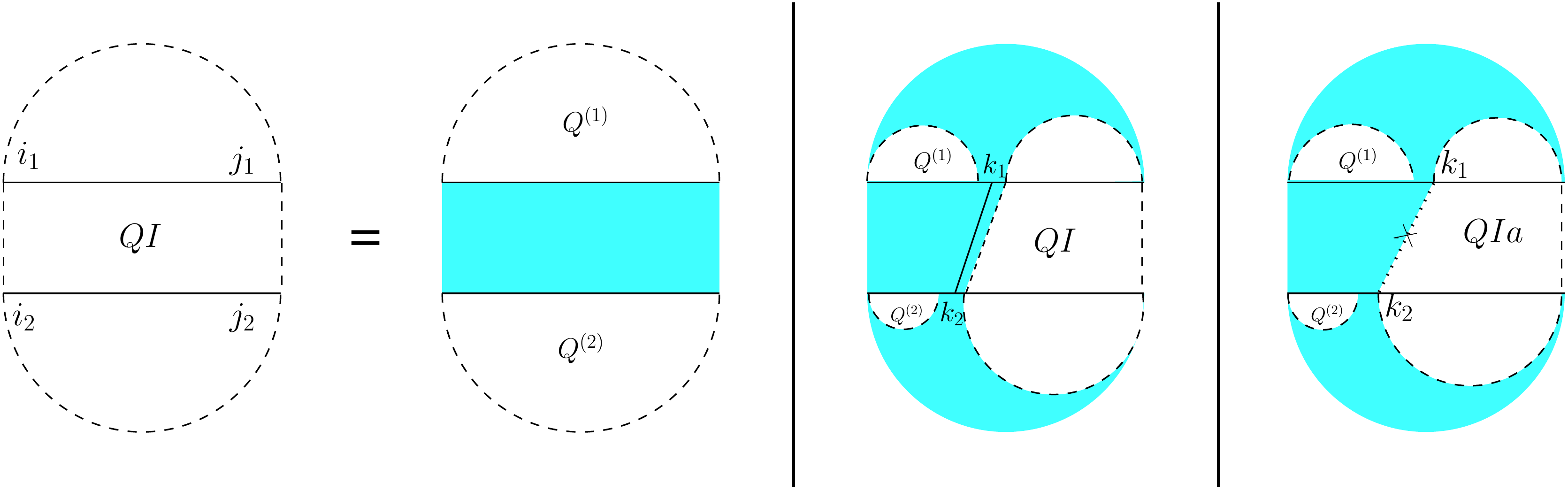}
\end{tabular}
\end{center}
\caption{Each case of a $\mathrm{QI}$ structure (left side of the equation)
  can lead to three cases: either no bonds exist (leftmost case), or at least
  one bond exists.  If the first event on both of the sequences is a bond
  (middle case) the subsequences to the left of the bond involve only
  $\mathrm{Q}$ and the subsequences to the right recurs on $\mathrm{QI}$.
  Otherwise (rightmost case) we will have $\mathrm{QIa}$ (see
  Figure~\ref{fig:QIa}).}
\label{fig:QI}
\end{figure*}

\begin{equation}
  \label{equ_Q}
  \mathrm{Q}_{i,j}= \left\{
    \begin{array}{ll}
      1 & j\leq i \\
      1+\displaystyle\sum_{k=i}^{j-4}\mathrm{Qz}_{k,j} &\mathrm{otherwise,}
    \end{array} \right. 
\end{equation}
\begin{equation}
  \label{equ_Qz}
  \mathrm{Qz}_{i,j}= \left\{
    \begin{array}{ll}
      0 & j-i < 4 \\
      \displaystyle \sum_{k=i+4}^{j} \mathrm{Q}_{i+1,k-1} \times e^{\mathrm{score}(i,k)} \times \mathrm{Q}_{k+1,j}  
        & \mathrm{otherwise.}\\
    \end{array}\right. 
\end{equation}

For the partition function of a pair of RNA sequences, we consider a
$4$-dimensional table $\mathrm{QI}$ in which $\mathrm{QI}_{i_1,j_1,i_2,j_2}$
is the value of base pair counting partition function for the subsequences
$[i_1,j_1]$ on $\RR$ and $[i_2,j_2]$ on $\RS$.  As Figure~\ref{fig:QI} shows,
we can split the set of all possible structures of $\mathrm{QI}$ into three
mutually exclusive subsets.  The leftmost case shows the structures in which
there exist no bonds (the first term of Equation~(\ref{equ_QI}).
The other two cases occur when there is at least one bond, and hence, unique
leftmost events on both $\RR$ and $\RS$, at positions $k_1$ and $k_2$,
respectively.  In the second (middle) case, these leftmost events are
end points of a bond, $k_1 \circ k_2$; hence, this case can be broken into: a
bond-free section on the left of the bond itself, and a general case of
$\mathrm{QI}$ on the right of the bond.  The third case occurs when $k_1$ and
$k_2$ are not end points of a bond.  We call this structure $\mathrm{QIa}$,
and explain it next.

\begin{eqnarray}
  \lefteqn{\mathrm{QI}_{i_1,j_1,i_2,j_2} =} \nonumber \\
  &&  \mathrm{Q}_{i_1,j_1}^{(1)} \times \mathrm{Q}_{i_2,j_2}^{(2)} + \nonumber \\
  && \displaystyle\sum_{k_1=i_1}^{j_1} \displaystyle\sum_{k_2=i_2}^{j_2} L_{i_1,j_1,k_1,i_2,j_2,k_2} + \nonumber \\
  && \displaystyle\sum_{k_1=i_1}^{j_1}
     \displaystyle\sum_{k_2=i_2}^{j_2} \left( \mathrm{Q}_{i_1,k_1-1}^{(1)}
     \times \mathrm{Q}_{i_2,k_2-1}^{(2)} \times \mathrm{QIa}_{k_1, j_1, k_2,
     j_2} \right),     
     \label{equ_QI}
\end{eqnarray}

\begin{equation}
L_{i_1,j_1,k_1,i_2,j_2,k_2} = 
 \mathrm{Q}_{i_1,k_1-1}^{(1)} \times \mathrm{Q}_{i_2,k_2-1}^{(2)} \times
     e^{\mathrm{iscore}(k_1,k_2)} \times \mathrm{QI}_{k_1+1, j_1, k_2+1, j_2}. 
\label{helper_2}
\end{equation}

\begin{figure*}[t]
\begin{center}
\begin{tabular}{cc}
\includegraphics[width=0.9\textwidth]{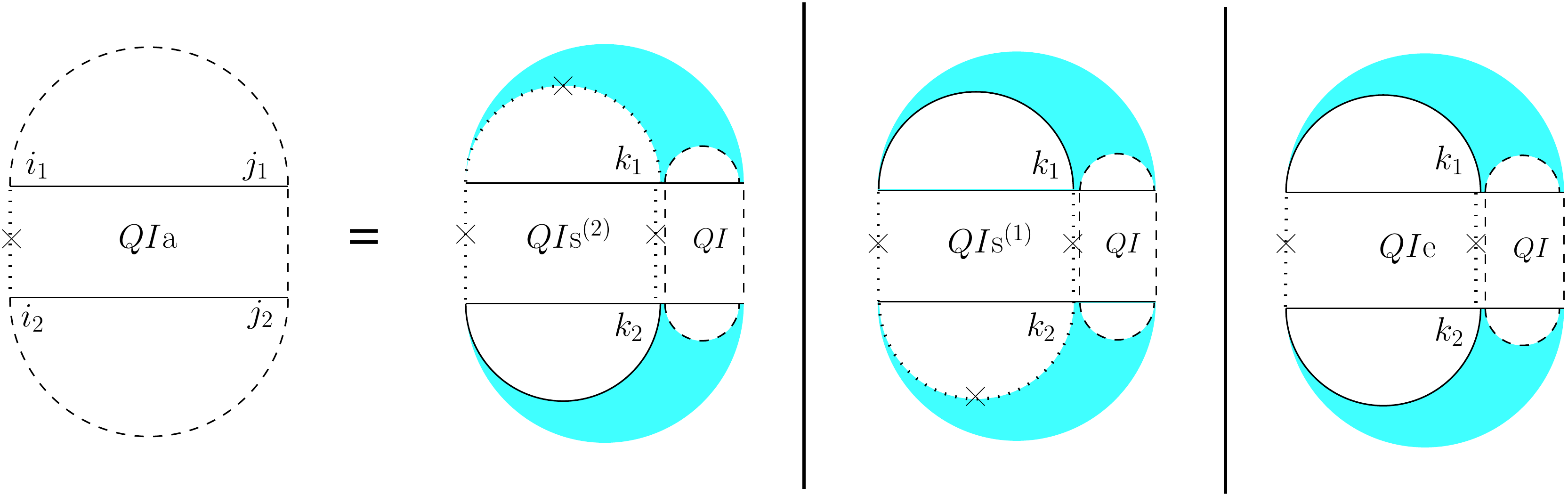}
\end{tabular}
\end{center}
\caption{There are three cases for computing the $\mathrm{QIa}$ structure;
  either the leftmost base of only one of the strands is an end point of an
  arc or both end points are.}
\label{fig:QIa}
\end{figure*}

For computing $\mathrm{QIa}_{i_1,j_1,i_2,j_2}$, (see Figure~\ref{fig:QIa}) we
have to consider the property of this structure that the leftmost bases on
both $\RR$ and $\RS$ have to be events, but they cannot both be the end points
of a bond.  Therefore, either one or both of them have to be end points of an
interaction arc.  There are two possibilities.

First, if both $i_1$ and $i_2$ are end points of some interaction arcs,
$i_1 \bullet k_1$ and $i_2 \bullet k_2$, these arcs must be equivalent (or
else, we have a zigzag).  As shown in the rightmost diagram in
Figure~\ref{fig:QIa}, $\mathrm{QIa}$ then splits into two exclusive
substructures, namely one where the first and last bases on each strand are
paired, and the two arcs are equivalent (we call it
$\mathrm{QIe}_{i_1,k_1,i_2,k_2}$ and derive its recursion later), followed by
$\mathrm{QI}_{k_1+1,j_1,k_2+1,j_2}$ on the suffixes of these arcs.

Otherwise, exactly one of the leftmost events on $\RR$ and $\RS$ is an
end point of a bond, and we have two symmetric cases ($\mathrm{QIs}^{(1)}$ and
$\mathrm{QIs}^{(2)}$), one where the interaction arc is in \RR, and the other
where it is in \RS.  In the first case (middle diagram in
Figure~\ref{fig:QIa}), let $k_1$ be the event in \RR\ such that
$i_1\bullet k_1$ is an interaction arc, and $[i_2,k_2]$ is the longest
subsequence in \RS\ that $i_1\bullet k_1$ subsumes, and $k_2$ is an event.
The suffix of this substructure recurs on $\mathrm{QI}$.  We derive
$\mathrm{QIs}^{(1)}$ later.

To derive $\mathrm{QIe}_{i_1,j_1,i_2,j_2}$, note that removing the arcs
$i_1 \bullet j_1$ and $i_2 \bullet j_2$ yields the general case of
$\mathrm{QI}_{i_1+1,j_1-1,i_2+1,j_2-1}$ for the inner-section with an
additional constraint that there has be at least one bond in that region
because the assumption is that the extracted arcs were interaction arcs.  We
can fulfill this constraint by excluding all cases where no bonds exist (i.e.,
considering only the two rightmost substructures of Figure~\ref{fig:QI}).

\begin{figure*}[t]
\begin{center}
\begin{tabular}{cc}
\includegraphics[width=0.5\textwidth]{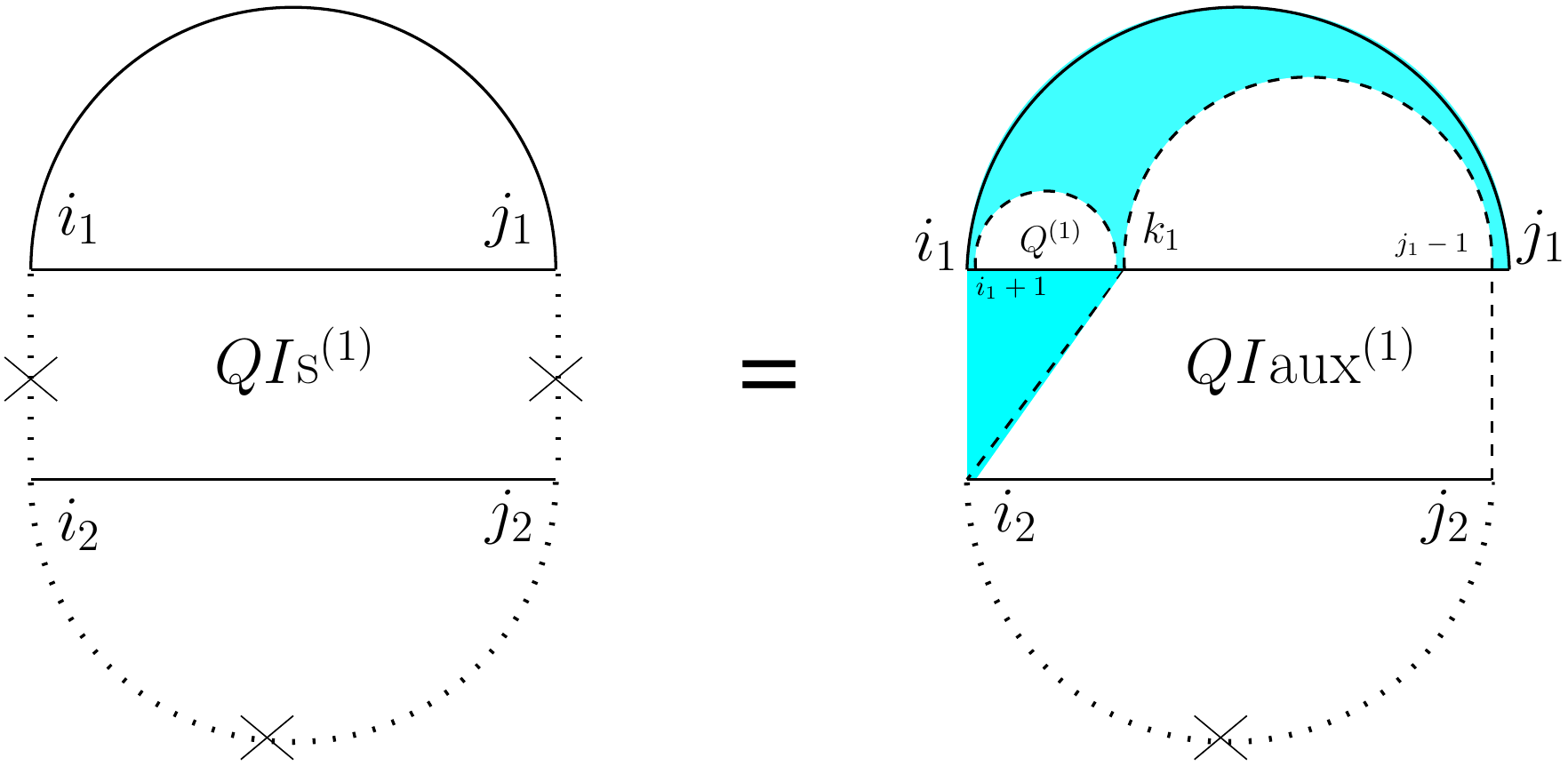}
\end{tabular}
\end{center}
\caption{$\mathrm{QIs}^{(1)}$ has one arc that can be extracted and the structure
  derived will have the property that the two end bases of the bottom strand
  cannot be paired (the new structure inherits this property from
  $\mathrm{QIs}^{(1)}$).  On the top strand, we consider the leftmost event.  This new
  structure is $\mathrm{QIaux}^{(1)}$.}
\label{fig:QIs}
\end{figure*}

To derive $QIs^{(1)}_{i_1,j_1,i_2,j_2}$ let $k_1$ be the leftmost event in the
subsequence $[i_1+1,j_1-1]$.  Note that such a $k_1$ is guaranteed to exist
because first, $i_1 \bullet j_1$ subsumes $[i_2,j_2]$ and we know that $i_2$
is an event, i.e., the end point of either a bond (subsumed by
$i_1\bullet j_1$) or of an interaction arc.  Then (see Figure~\ref{fig:QIs})
we define a new substructure, $\mathrm{QIaux}^{(1)}$, after removing
$i_1 \bullet j_1$ and the prefix of \RR\ up to $k_1$.

\begin{figure*}[t]
\begin{center}
\begin{tabular}{cc}
\includegraphics[width=0.75\textwidth]{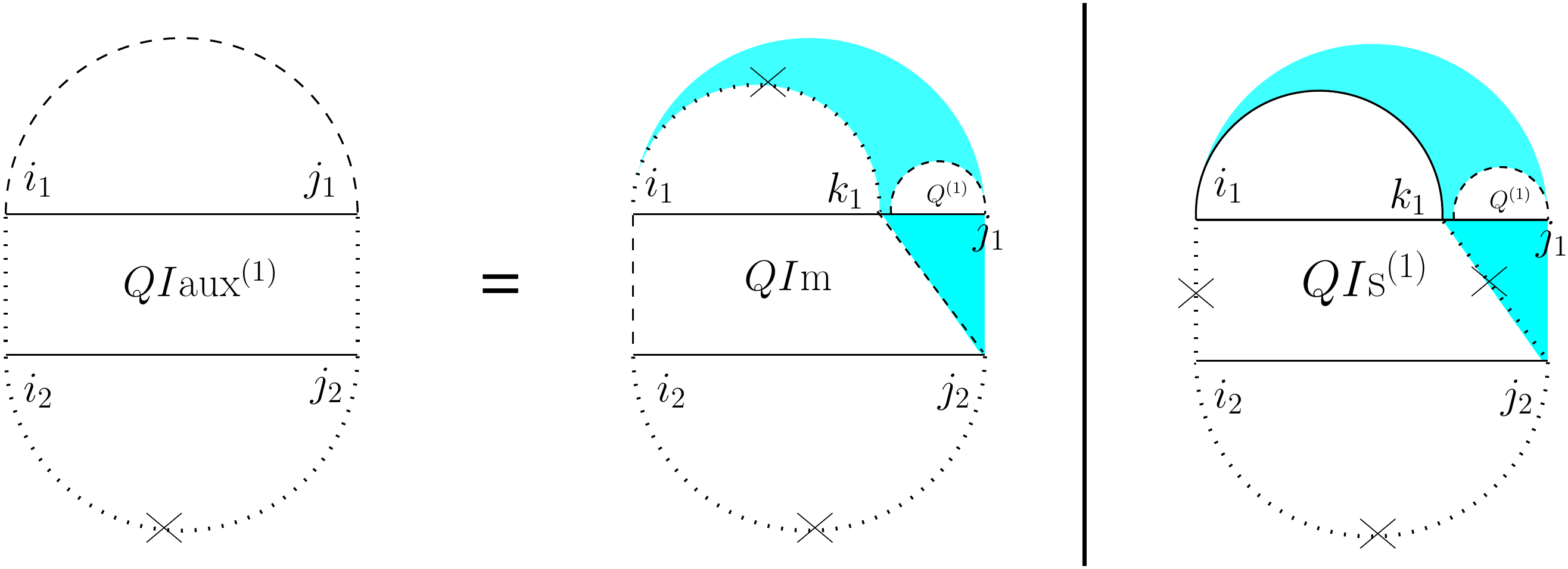}
\end{tabular}
\end{center}
\caption{Two cases must be considered for the $\mathrm{QIaux}^{(1)}$
  structure, in which the two end points of the bottom strand are events.  For
  the top strand, only the leftmost end point is required to be an event.  It
  can either be the end point of an arc (rightmost case) or not (leftmost
  case).}
\label{fig:QIaux}
\end{figure*}

To derive $\mathrm{QIaux}^{(1)}_{i_1,j_1,i_2,j_2}$, note that the context of
its definition implies that $i_1, i_2$ and $j_2$ are all three events.  Let,
as shown in Figure~\ref{fig:QIaux}, $k_1$ be the \emph{last} event on
$[i_1,j_1]$.  Now, if $i_1\bullet k_1$, then recur on $\mathrm{QIs}^{(1)}$.  Otherwise, $k_1$ is an event that does not pair with $i_1$.
We define a new substructure, $\mathrm{QIm}$, where all four corners are
events, and neither $i_1 \bullet j_1$ nor $i_2 \bullet j_2$ is allowed.

\begin{figure*}[t]
\begin{center}
\begin{tabular}{cc}
\includegraphics[width=0.94\textwidth]{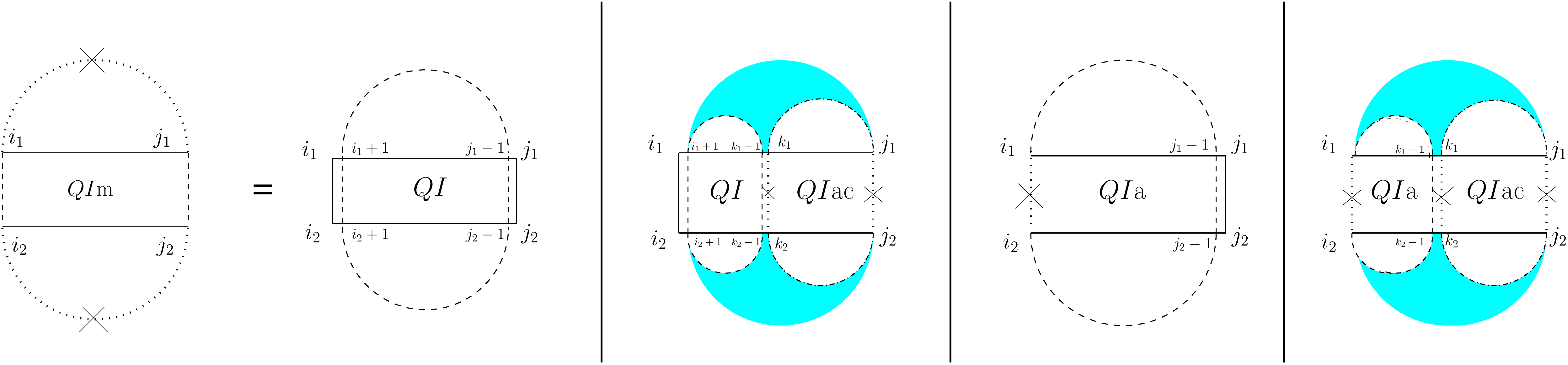}
\end{tabular}
\end{center}
\caption{For computing $\mathrm{QIm}$, since we know the four end points are
  events, but none of the two end points in one strand can form an arc, we
  must consider the four different cases shown above. For convenience, arcs of $\mathrm{QIac}$ structure are shown with dash-dotted lines because it represents the sum of three structures in which each of the arcs could be present or not (we could replace the second and fourth cases with three cases, one for each term of Equation~(\ref{equ_QIac})). }
\label{fig:QIm}
\end{figure*}

For computing $\mathrm{QIm}_{i_1,j_1,i_2,j_2}$, since there are four corners each of which can be the end point of either a bond or of an arc, there might be at most sixteen possibilities. Upon combining some of those sixteen possibilities, we have to consider four
mutually exclusive cases (see Figure~\ref{fig:QIm}). The first one is the
case where $i_1 \circ i_2$ and $j_1 \circ j_2$ and the remaining part will be
$\mathrm{QI}_{i_1+1,j_1-1,i_2+1,j_2-1}$. That case corresponds to all four corner events being the end points of bonds. Since we assume there are no crossing bonds, we must have $i_1 \circ i_2$ and $j_1 \circ j_2$. In the second case, $i_1$ and $i_2$ are the end points of a bond, i.e., $i_1 \circ i_2$,
but either $j_1$ or $j_2$ (or both) does not form a bond. That captures three out of the sixteen total possibilities. Since $j_1$ and $j_2$ are both events
but do not form a bond, we define a term $\mathrm{QIac}$ which is the sum of
$\mathrm{QIe}$ and the two symmetric $\mathrm{QIs}$'s, since they preserve the
constraints that arise in the first term in the definition of $\mathrm{QIa}$
(see Figure~\ref{fig:QIa}).  The prefix of this substructure is a general
recursion on $\mathrm{QI}$ on the subsequences $[i_1+1, k_1-1]$ and
$[i_2+1, k_2-1]$.  The third case is the symmetric case of the second case, i.e., there is no bond between $i_1$ and
$i_2$, but $j_1 \circ j_2$.  The prefix of this bond is a recursion on
$\mathrm{QIa}$. That captures three out of the sixteen total possibilities. Finally, the fourth case corresponds to either $i_1$ or $i_2$ (or both) does not form a bond and either $j_1$ or $j_2$ (or both) does not form a bond. That captures the remaining nine out of the sixteen total possibilities.

Putting all those together, we obtain
\begin{equation}
\label{equ_QIa}
\mathrm{QIa}_{i_1,j_1,i_2,j_2}= 
 \displaystyle\sum_{k_1=i_1}^{j_1} \displaystyle\sum_{k_2=i_2}^{j_2} 				\mathrm{QIac}_{i_1,k_1,i_2,k_2} \times \mathrm{QI}_{k_1+1,j_1,k_2+1,j_2},
\end{equation} 

\begin{equation}
\label{equ_QIac}
\mathrm{QIac}_{i_1,j_1,i_2,j_2} = 
 \mathrm{QIs}_{i_1,j_1,i_2,j_2}^{(1)} + \mathrm{QIs}_{i_1,j_1,i_2,j_2}^{(2)} + \mathrm{QIe}_{i_1,j_1,i_2,j_2},
\end{equation}

\begin{equation}
\mathrm{QIe}_{i_1,j_1,i_2,j_2}= 
    \left\{
    \begin{array}{ll}
      0 & j_1 - i_1 < 4 \\
        & \mathrm{or\ \ } j_2 - i_2 < 4 \vspace*{3mm} \\
      M_{i_1,j_1,i_2,j_2}
        & \mathrm{otherwise,}
    \end{array} \right.
  \label{equ_QIe}
\end{equation}

\begin{equation}
M_{i_1,j_1,i_2,j_2} = 
 \left(\mathrm{QI}_{i_1+1,j_1-1,i_2+1,j_2-1} - \mathrm{Q}_{i_1+1,j_1-1}^{(1)}
      \times \mathrm{Q}_{i_2+1,j_2-1}^{(2)}\right) 
      \times e^{\mathrm{score}(i_1,j_1) + \mathrm{score}(i_2,j_2)},
\label{helper_3}
\end{equation}

\begin{equation}
\label{equ_QIs1}
\mathrm{QIs}_{i_1,j_1,i_2,j_2}^{(1)}= 
\left\{
\begin{array}{ll}
  0 & j_1 - i_1 < 4 \mathrm{\ \ or\ \ } j_2 < i_2  \\ \\
  \displaystyle\sum_{k_1=i_1+1}^{j_1-1}
  \mathrm{Q}_{i_1+1,k_1-1}^{(1)} \times e^{\mathrm{score}(i_1,j_1)} \times
  \mathrm{QIaux}_{k_1, j_1-1, i_2, j_2}^{(1)}
    & \mathrm{otherwise,}\\
\end{array} \right.
\end{equation}

\begin{equation}
\label{equ_QIs2}
\mathrm{QIs}_{i_1,j_1,i_2,j_2}^{(2)}= 
\left\{
\begin{array}{ll}
  0 & j_1 < i_1 \mathrm{\ \ or\ \ } j_2 - i_2 < 4 \\ \\
  \displaystyle\sum_{k_2=i_2+1}^{j_2-1}
  \mathrm{Q}_{i_2+1,k_2-1}^{(2)} \times e^{\mathrm{score}(i_2,j_2)} \times
   \mathrm{QIaux}_{i_1, j_1, k_2, j_2-1}^{(2)}
    & \mathrm{otherwise,}
\end{array} \right.
\end{equation}

\begin{equation}
\label{equ_QIaux1}
\mathrm{QIaux}_{i_1,j_1,i_2,j_2}^{(1)} =    
    \displaystyle\sum_{k_1=i_1}^{j_1}
    \left(\mathrm{QIs}_{i_1,k_1,i_2,j_2}^{(1)} + \mathrm{QIm}_{i_1,k_1,i_2,j_2}\right) \times \mathrm{Q}_{k_1+1,j_1}^{(1)},  
\end{equation}

\begin{equation}
\label{equ_QIaux2}
\mathrm{QIaux}_{i_1,j_1,i_2,j_2}^{(2)} =    
    \displaystyle\sum_{k_2=i_2}^{j_2}
    \left(\mathrm{QIs}_{i_1,j_1,i_2,k_2}^{(2)} + \mathrm{QIm}_{i_1,j_1,i_2,k_2}\right) \times \mathrm{Q}_{k_2+1,j_2}^{(2)},  
\end{equation}

\begin{equation}
\mathrm{QIm}_{i_1,j_1,i_2,j_2}=
\left\{
\begin{array}{ll}
  e^{\mathrm{iscore}(i_1,i_2)} & i_1=j_1 \mathrm{\ \ and\ \ } i_2=j_2 \vspace*{3mm} \\
  N_{i_1,j_1,i_2,j_2} & i_1 < j_1 \mathrm{\ \ and\ \ } i_2 < j_2 \vspace*{3mm} \\
  0 & \mathrm{otherwise,}
    \end{array} \right.
  \label{equ_QIm}
\end{equation}

\begin{equation}
\begin{array}{lll}
    \lefteqn{N_{i_1,j_1,i_2,j_2} = } && \\ [4mm]
    && e^{\mathrm{iscore}(i_1,i_2)+\mathrm{iscore}(j_1,j_2)} \times \mathrm{QI}_{i_1+1,j_1-1,i_2+1,j_2-1} + \\
  && e^{\mathrm{iscore}(i_1,i_2)}  \times  
  \displaystyle\sum_{k_1=i_1+1}^{j_1}
  \displaystyle\sum_{k_2=i_2+1}^{j_2}
  \mathrm{QI}_{i_1+1,k_1-1,i_2+1,k_2-1}
  \times \mathrm{QIac}_{k_1,j_1,k_2,j_2} + \\
  && e^{\mathrm{iscore}(j_1,j_2)} \times \mathrm{QIa}_{i_1,j_1-1,i_2,j_2-1} + \\
  && \displaystyle\sum_{k_1=i_1}^{j_1}
  \displaystyle\sum_{k_2=i_2}^{j_2}
  \mathrm{QIa}_{i_1,k_1,i_2,k_2}
  \times \mathrm{QIac}_{k_1+1,j_1,k_2+1,j_2}.
\end{array}
\label{helper_4}
\end{equation}

\section{Results}
\label{sec:results}
\begin{figure*}[t]
\begin{center}
\begin{tabular}{cc}
\includegraphics[angle=-90,width=0.47\textwidth]{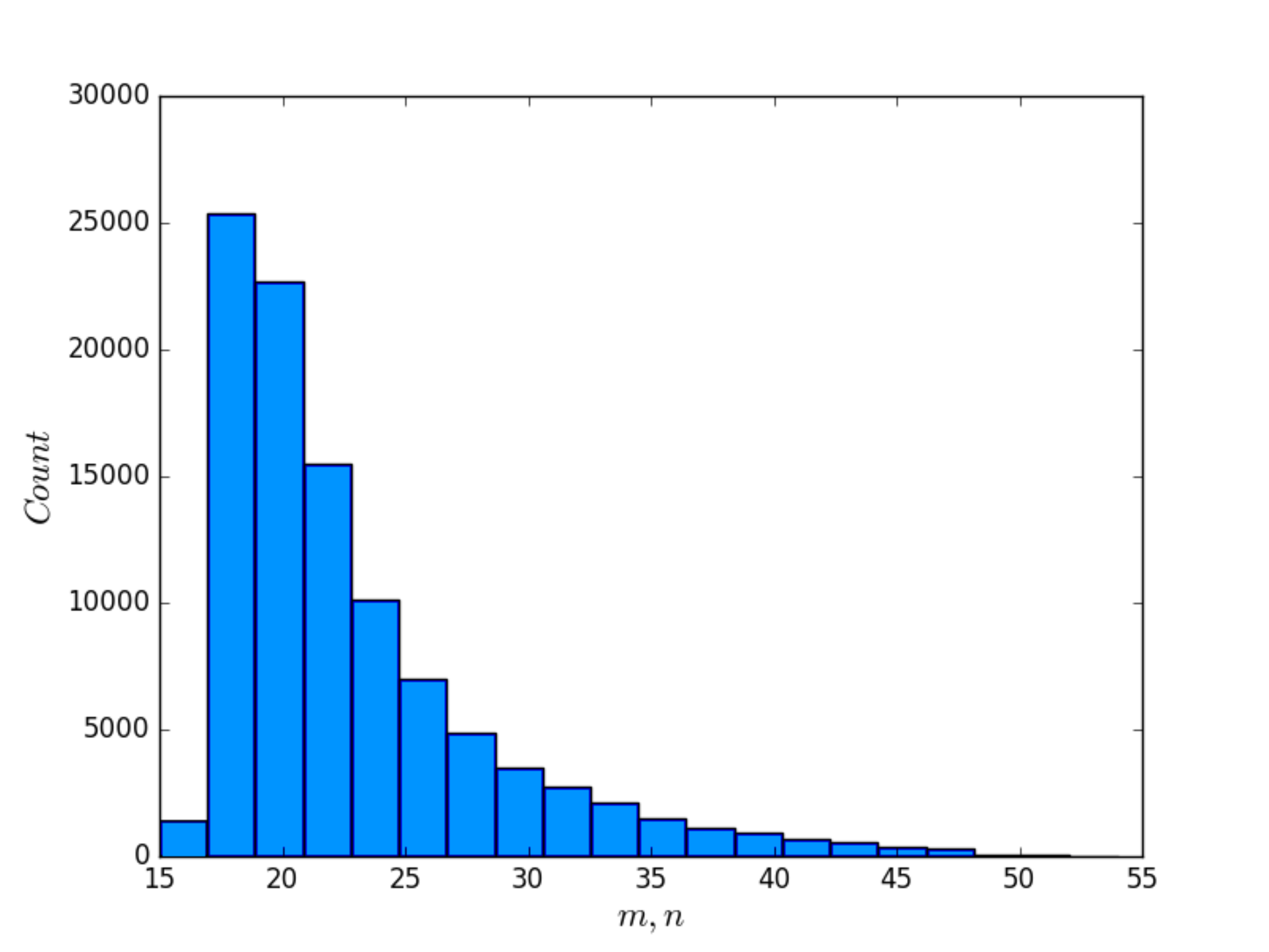}&
\includegraphics[angle=-90,width=0.47\textwidth]{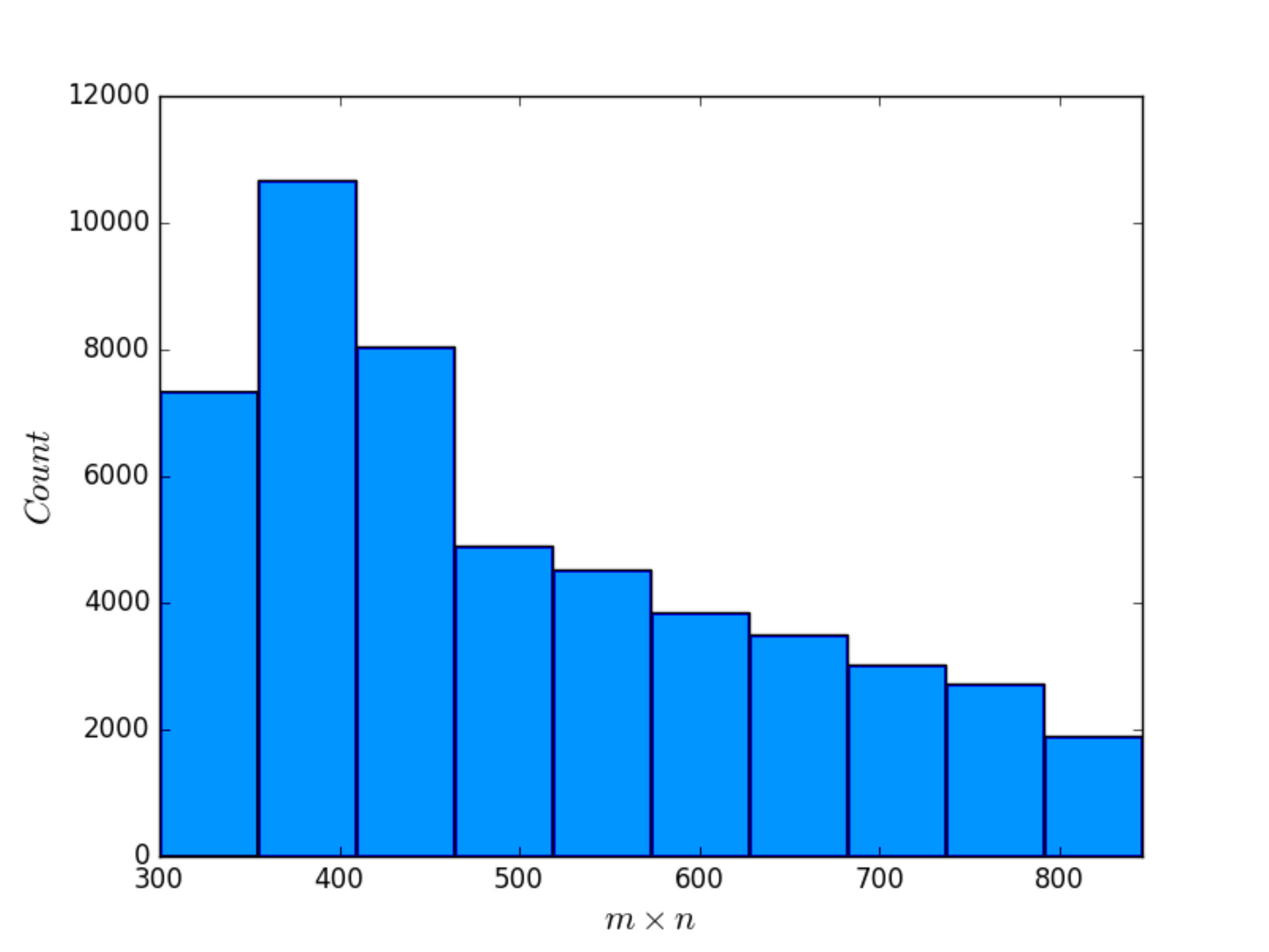}\\
\end{tabular}
\end{center}
\caption{ Distribution of the lengths (left) and product of the lengths (right) of all the RNA subsequences used in our
  experiment.}\label{fig:dist}
\end{figure*}

To investigate the correlation between the scores of \texttt{BPPart} and
\texttt{BPMax}, and those of \texttt{piRNA}, we used the RISE database~\cite{Gong17} which
combines information about interacting RNAs from multiple experiments.  For
the human dataset, we extracted all the interaction windows for those pairs that
have this data in RISE.  We eliminated the ones that contained a window with length
less than $15$ because they are too short for an unbiased comparison.  Then,
we sorted the remaining pairs based on the product of the lengths of the
interacting windows.  Finally, the first 50,500~pairs of sequences were chosen
as our primary dataset for different experiments and analysis.
Figure~\ref{fig:dist} shows the distribution of the sequence lengths in this
dataset, and also the product of the lengths of the RNA subsequences in each
pair.

We first ran \texttt{piRNA} on our primary dataset at 8 different
temperatures, $37$, $25$, $13$, $0$, $-40$, $-80$, $-130$, and $-180$ degrees
Celsius.  We also ran \texttt{BPPart} and \texttt{BPMax} on the dataset with
different weights, i.e., $c_i$'s and $c'_i$'s.  In general, we want to use
the stack energies of the Turner model as a starting point for tuning the
weights.  Since the parameters form a projective space (invariant results with respect to scaling), we considered a fixed weight of $3$ for \texttt{CG} (and
\texttt{GC}). Using the experimentally computed stack energies of the Turner
model, minimum and maximum values for the weights of \texttt{AU} and
\texttt{GU} were computed. That is, to compute the maximum weight of \texttt{AU}
(and \texttt{UA}), we consider the maximum released energy when \texttt{AU}
(or \texttt{UA}) is stacked with another pair; this happens when \texttt{UA}
is stacked with \texttt{CG} and $2.4\;kcal/mol$ energy is released.  Then, we
considered the minimum value of released energy in an stack for \texttt{CG} or
\texttt{GC} (for which we assumed a constant weight of $3$), which is
$1.4\;kcal/mol$.  We derived the maximum weight of \texttt{AU} and \texttt{UA}
as $5.143$ by multiplying $2.4$ by $\frac{3}{1.4}$.  Finally, we made sure
that the range of values that we explore for the weight of \texttt{AU} and
\texttt{UA} contains this maximum value (we chose $5.5$ as the upper-bound).
For finding the minimum weight of \texttt{AU} and \texttt{UA}, we consider
their minimum stack energy, which is $0.6\;kcal/mol$.  Given the maximum
energy of \texttt{CG}, namely $3.4\;kcal/mol$, the value of interest is
computed as $0.6\times \frac{3}{3.4} = 0.529$.  However, for the sake of
comprehensiveness and exploring the shape of the plots, we used much smaller
lower-bounds---$-4.5$ and $-3$---for \texttt{BPPart} and \texttt{BPMax},
respectively.

\begin{figure*}[t]
\begin{center}
\begin{tabular}{cc}
\includegraphics[angle=-90,width=0.45\textwidth]{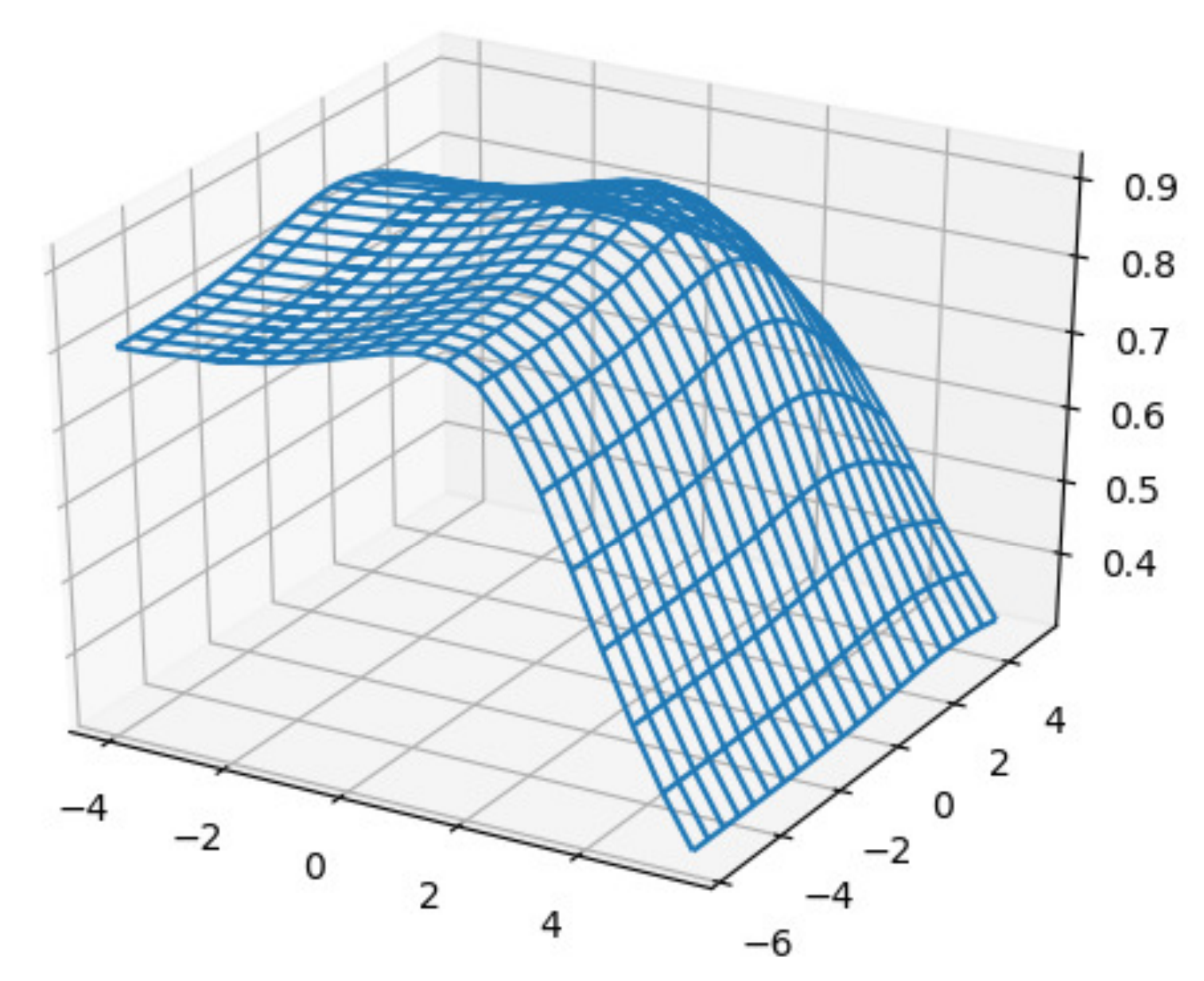}&
\includegraphics[angle=-90,width=0.45\textwidth]{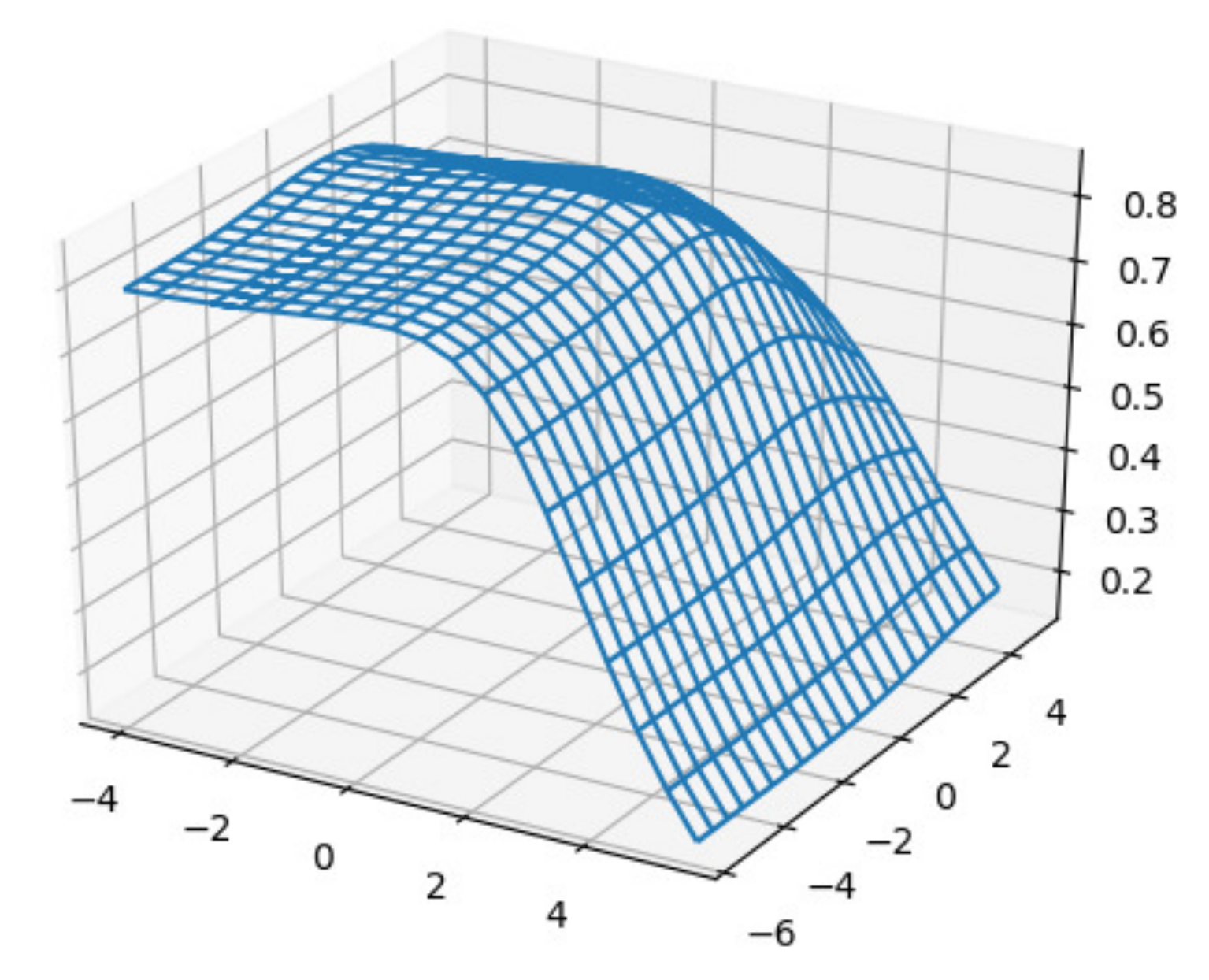}\\
\end{tabular}
\end{center}
\caption{Pearson correlation between \texttt{piRNA} and \texttt{BPPart}
  (vertical axis), on the primary dataset, at $-180^{\circ}C$ (left) and
  $37^{\circ}C$ (right) for different values of constant factors (weights) for
  $AU$ (left axis) and $GU$ (right axis).  The weight of $CG$ pair is fixed at
  $3$.}
\label{fig:temp-cor_1}
\end{figure*}

\begin{figure*}[t]
\begin{center}
\begin{tabular}{cc}
\includegraphics[angle=-90,width=0.45\textwidth]{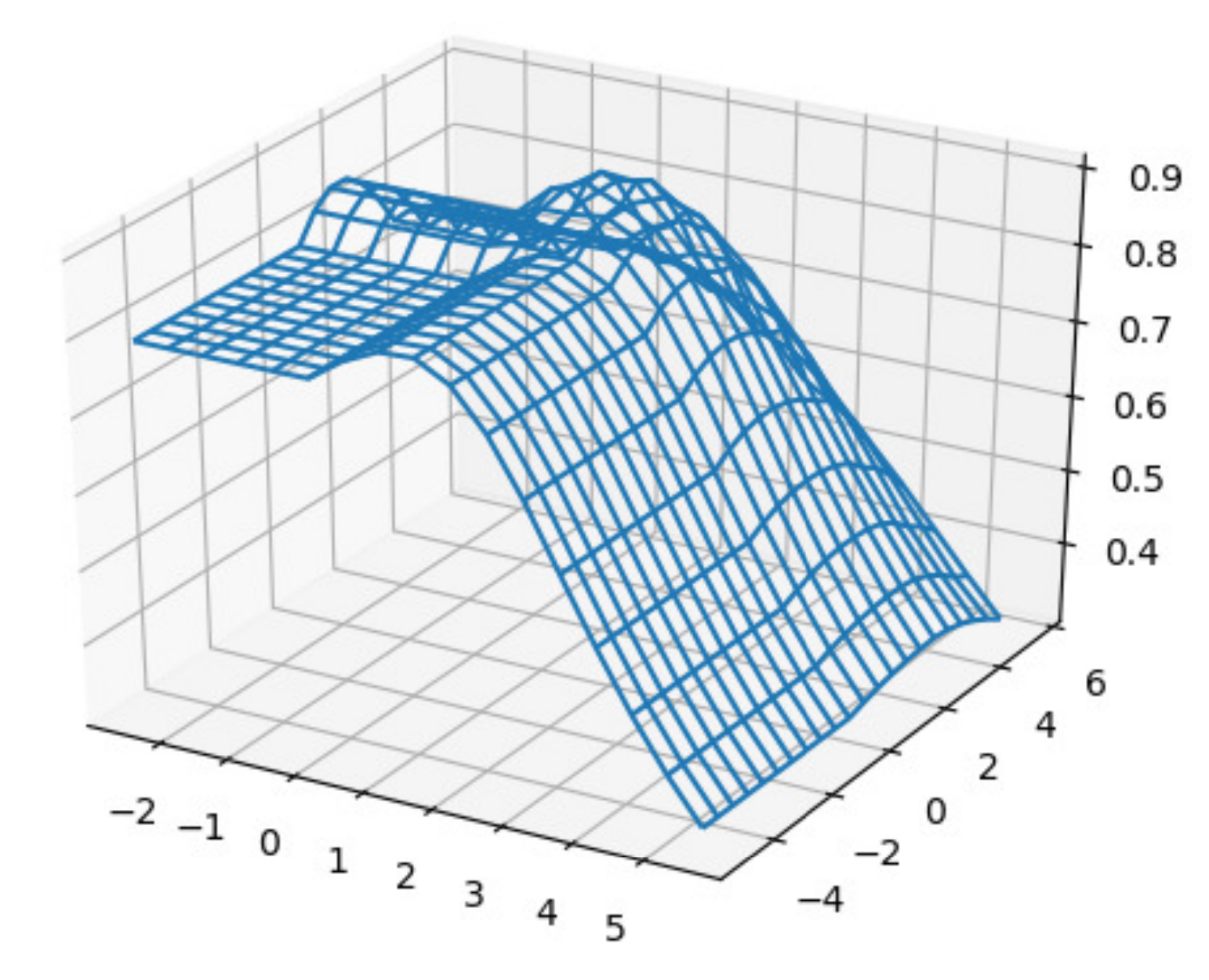}&
\includegraphics[angle=-90,width=0.45\textwidth]{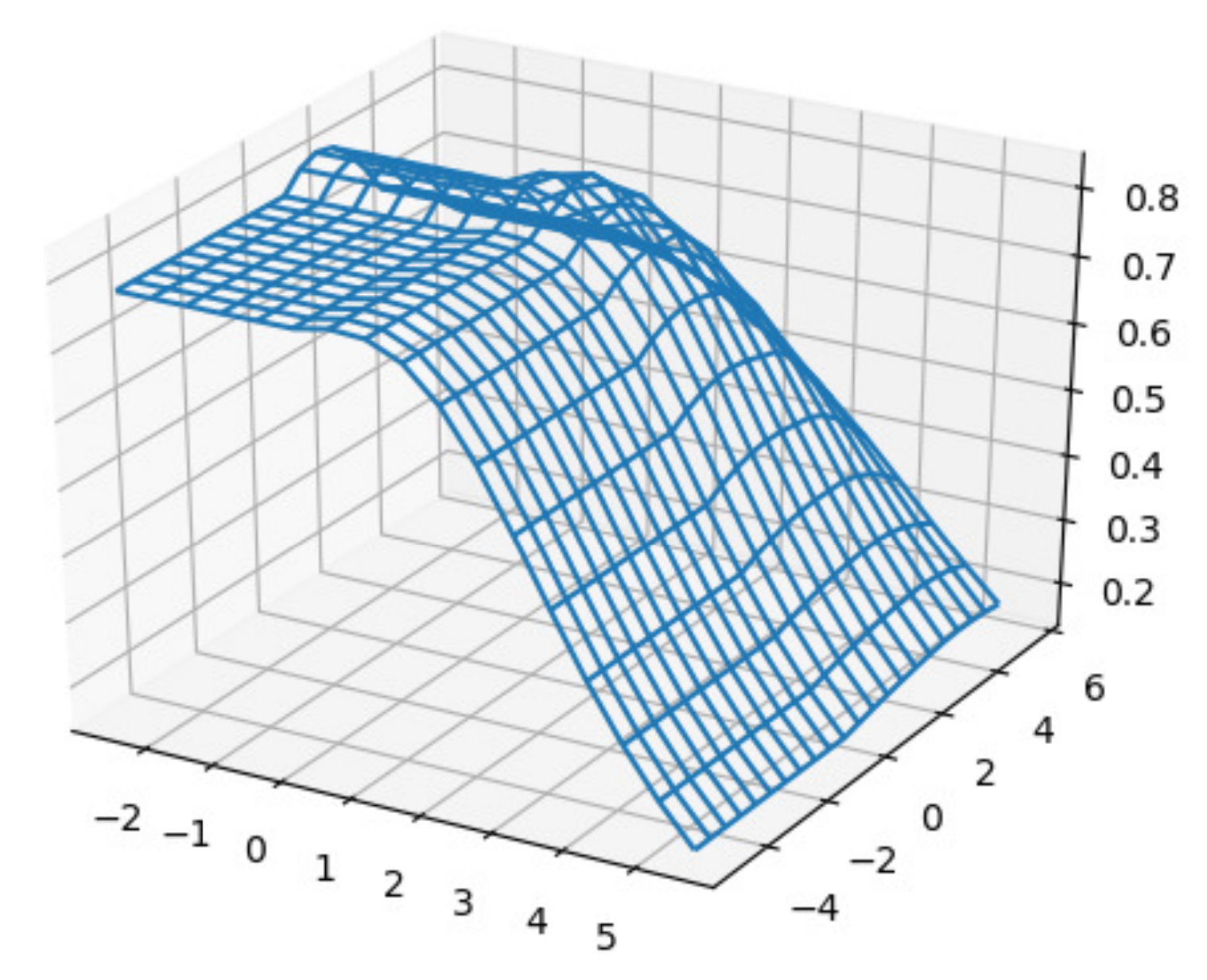}\\
\end{tabular}
\end{center}
\caption{Pearson correlation between \texttt{piRNA} and \texttt{BPMax}
  (vertical axis), on the primary dataset, at $-180^{\circ}C$ (left) and
  $37^{\circ}C$ (right) for different values of constant factors (weights) for
  $AU$ (left axis) and $GU$ (right axis).  The weight of $CG$ pair is fixed at
  $3$.}
\label{fig:temp-cor_2}
\end{figure*}

\begin{figure*}[t]
\begin{center}
\begin{tabular}{cc}
\includegraphics[width=0.45\textwidth]{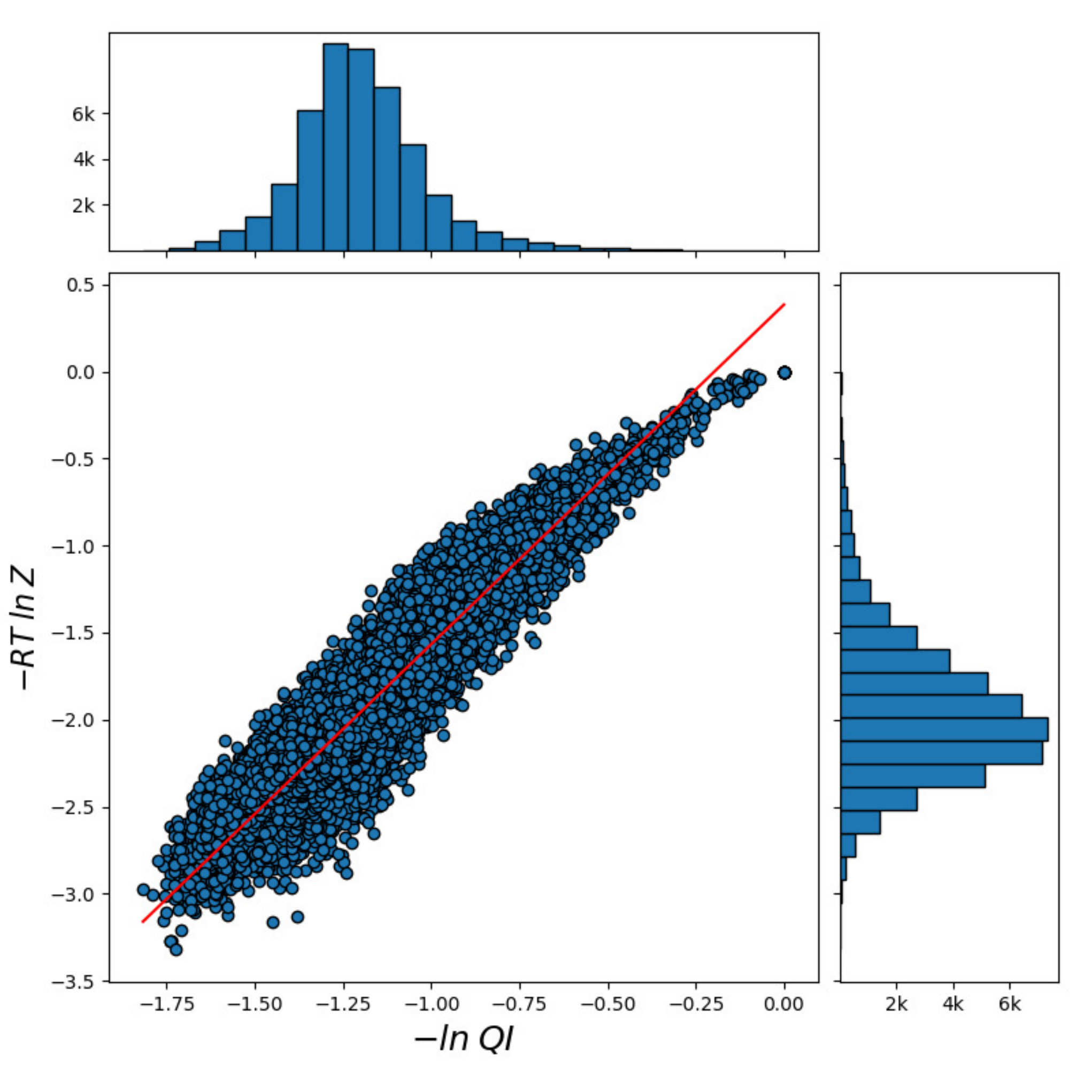}&
\includegraphics[width=0.45\textwidth]{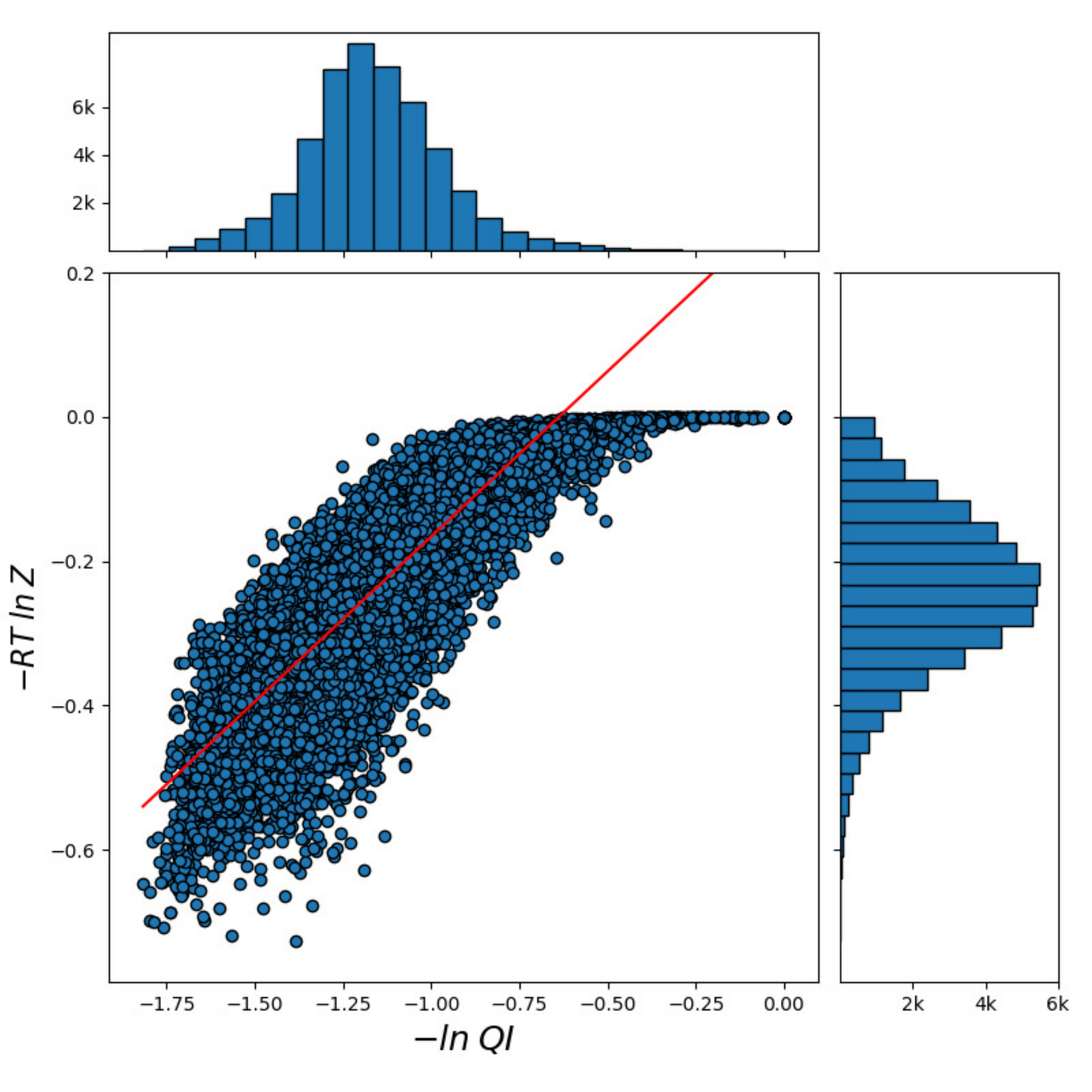}\\
\end{tabular}
\end{center}
\caption{Scatter plots of the values of the corresponding axis for each sample
  (interaction windows of a pair of RNAs from RISE dataset) at $-180^{\circ}C$
  (left) and $37^{\circ}C$ (right).  In both plots, the red line is a
  straight regression line fitted to the points by minimizing mean square error (MSE).}
\label{fig:cor1}
\end{figure*}

For all the combinations of weights of \texttt{AU} and \texttt{GU}, in steps
of $0.5$, we computed the Pearson and Spearman's rank correlations with the
scores from \texttt{piRNA} at different temperatures.  When computing the
correlations, to normalize the scores from all algorithms, we divide them
by the sum of the lengths of corresponding sequences, $L_R + L_S$.  This normalization
mitigates the effect of length bias on the computed correlations.  This step is
necessary because, generally, as the length of the pair of sequences increases
the scores of all three algorithms increases, and if unnormalized scores are
used, a biased higher correlation will be derived.  Note that for partition functions, \texttt{piRNA} and \texttt{BPPart}, we used the $log$
of the scores; that is why we factor out the sum of the lengths for
normalization.  If the original values were to be used, we would have to
take the $(L_R + L_S)$th roots of the scores. Figures~\ref{fig:temp-cor_1} and \ref{fig:temp-cor_2} show the Pearson correlations for different combinations of weights of \texttt{AU} and \texttt{GU} at $-180^{\circ}C$ and $37^{\circ}C$. Figure~\ref{fig:cor1} shows the scatter plots of the
scores of \texttt{BPPart} and \texttt{piRNA} at these temperatures. In these plots, the red line shows the regression line that is fitted to the
points by minimizing the mean squared error (MSE).

The optimum values of correlation for each temperature are presented in
Tables~\ref{cor:pearson} and \ref{cor:rank}. There is a high correlation
between \texttt{piRNA} and \texttt{BPPart} as well as between \texttt{piRNA}
and \texttt{BPMax}, especially when the temperature decreases which is due to
a decrease in the role of thermodynamic entropy at the lower temperatures.
Also, we computed the Pearson and Spearman's rank correlation between \texttt{BPPart} and
\texttt{BPMax} with their optimum weights at $37^{\circ}C$, which yielded values of 0.971 and 0.968, respectively.  

\begin{table*}[h]
\begin{center}
  \caption{Pearson correlation between \texttt{piRNA} and \texttt{BPPart} and
    between \texttt{piRNA} and \texttt{BPMax} at different temperatures ($T^{\circ}C$).}
  \begin{tabular}{||c | c c c c c c c c||} 
    \hline
    Method \textbackslash T & 37 & 25 & 13 & 0 & -40 & -80 & -130 & -180 \\ [0.5ex]
    \hline\hline
    BPPart & 0.855 & 0.862 & 0.869 & 0.877 & 0.896 & 0.908 & 0.916 & 0.920 \\ 
    \hline
    BPMax  & 0.836 & 0.846 & 0.855 & 0.864 & 0.884 & 0.895 & 0.901 & 0.904 \\ [1ex]
    \hline
  \end{tabular}
  \label{cor:pearson}
\end{center}
\end{table*} 

\begin{table*}[h]
  \begin{center}
    \caption{Spearman's rank correlation between \texttt{piRNA} and
      \texttt{BPPart} and between \texttt{piRNA} and \texttt{BPMax} at
      different temperatures ($T^{\circ}C$).}
    \begin{tabular}{||c | c c c c c c c c||} \hline
 Method \textbackslash T & 37 & 25 & 13 & 0 & -40 & -80 & -130 & -180
      \\[0.5ex] \hline\hline
      BPPart & 0.864 & 0.867 & 0.871 & 0.876 & 0.889 & 0.896 & 0.901 & 0.901
      \\\hline 
      BPMax  & 0.830 & 0.835 & 0.841 & 0.847 & 0.862 & 0.871 & 0.877 & 0.877
      \\[1ex] \hline
    \end{tabular}
    \label{cor:rank}
  \end{center}
\end{table*}

  

\begin{figure*}
\begin{center}
\begin{tabular}{cc}
\includegraphics[angle=0,width=0.45\textwidth]{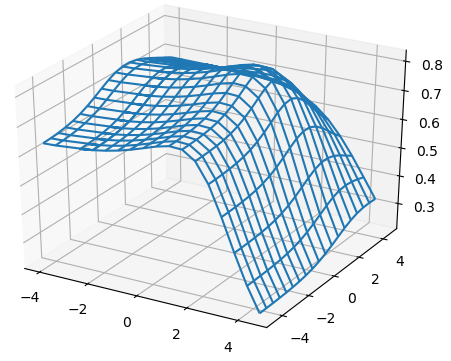}&
\includegraphics[angle=0,width=0.45\textwidth]{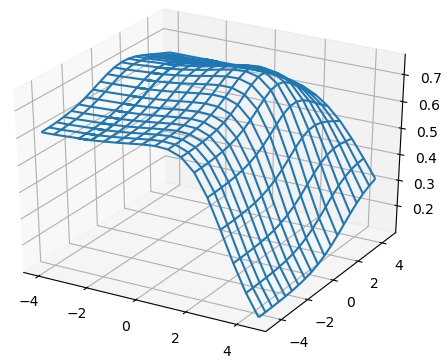}\\
\end{tabular}
\end{center}
\caption{Pearson correlation between \texttt{piRNA} and \texttt{BPPart}
  (vertical axis), on the randomly generated dataset, at $-180^{\circ}C$ (left) and
  $37^{\circ}C$ (right) for different values of constant factors (weights) for
  $AU$ (left axis) and $GU$ (right axis). The weight of $CG$ pair is fixed at
  $3$.}
\label{fig:rand-cor_1}
\end{figure*}

\begin{figure*}
\begin{center}
\begin{tabular}{cc}
\includegraphics[angle=0,width=0.45\textwidth]{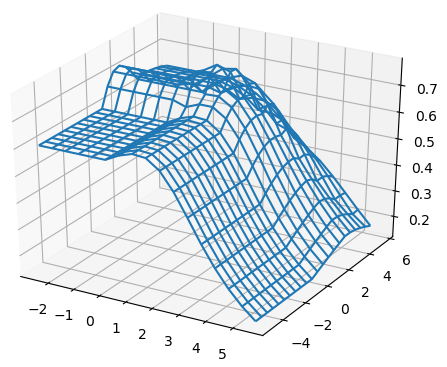}&
\includegraphics[angle=0,width=0.45\textwidth]{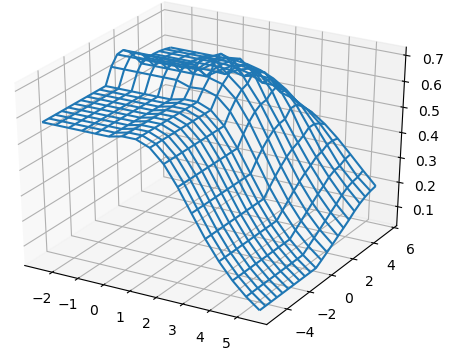}\\
\end{tabular}
\end{center}
\caption{Pearson correlation between \texttt{piRNA} and \texttt{BPMax}
  (vertical axis), on the randomly generated dataset, at $-180^{\circ}C$ (left) and
  $37^{\circ}C$ (right) for different values of constant factors (weights) for
  $AU$ (left axis) and $GU$ (right axis).  The weight of $CG$ pair is fixed at
  $3$.}
\label{fig:rand-cor_2}
\end{figure*}

To make sure that the optimization results are not data dependent, we
conducted the same experiments for randomly generated sequences.  To factor out
the effect of length, for each pair in our primary dataset, we generated a pair of random sequences with the same lengths as those of the pair in our primary dataset.  The shape of the plots are
very similar to those for the primary dataset, for both \texttt{BPPart} and
\texttt{BPMax} (see Figures~\ref{fig:rand-cor_1},~\ref{fig:rand-cor_2}).  For \texttt{BPPart}, the optimum
weights at $-180^{\circ}C$ are the same ($1.0$, $1.0$, and $3$ for \texttt{AU},
\texttt{GU}, and \texttt{CG}, respectively) and at $37^{\circ}C$ the optimum
weights we had earlier ($0.5$, $1.0$, and $3$) are ranked $3^\mathrm{rd}$ for
the random dataset with only $0.003$ difference in the Pearson correlation
from the optimum, achieved by weights of ($1.0$, $1.0$, and $3$) for the random dataset.
Similarly, for \texttt{BPMax}, the optimum values for the two datasets are the
same at $-180^{\circ}C$ ($1.0$, $2.0$, and $3$), and at $37^{\circ}C$, the
optimum values of weights for the primary dataset ($1.0$, $1.5$, and $3$) are
ranked $2^\mathrm{nd}$ for the random dataset with only $0.008$ difference in the
Pearson correlation from the optimum, achieved by weights of ($1.0$,
$2.0$, and $3$) for this dataset.


Although the shape of the plots and the peaks were almost the same for the
primary dataset and the random dataset, the best correlations for the random
one were considerably less than those of the primary one.
Table~\ref{cor:rand} shows the Pearson and the Spearman's rank correlation of
\texttt{BPPart} and \texttt{BPMax} with \texttt{piRNA} at $-180^{\circ}C$ and
$37^{\circ}C$ for the random dataset.

\begin{table*}[h]
  \begin{center}
    \caption{Pearson and Spearman's rank correlation between \texttt{piRNA}
      and \texttt{BPPart} and between \texttt{piRNA} and \texttt{BPMax} at
      $-180^{\circ}C$ and $37^{\circ}C$ for the random input sequences.}
\begin{tabular}{c | c c | c c  }
 &
  \multicolumn{2}{c}{Pearson} &
  \multicolumn{2}{c}{Spearman} \\
\cline{2-3}
\cline{4-5}
  Method \textbackslash T  & $37^{\circ}C$ & $-180^{\circ}C$ & $37^{\circ}C$ & $-180^{\circ}C$ \\
\hline
\hline
BPPart      & 0.761 & 0.825 & 0.753 & 0.801  \\
BPMax       & 0.716 & 0.785 & 0.702 & 0.751  \\

\end{tabular}
 \label{cor:rand}
\end{center}
\end{table*}

\begin{figure*}[t]
\begin{center}
\begin{tabular}{cc}
\includegraphics[angle=-90,width=0.45\textwidth]{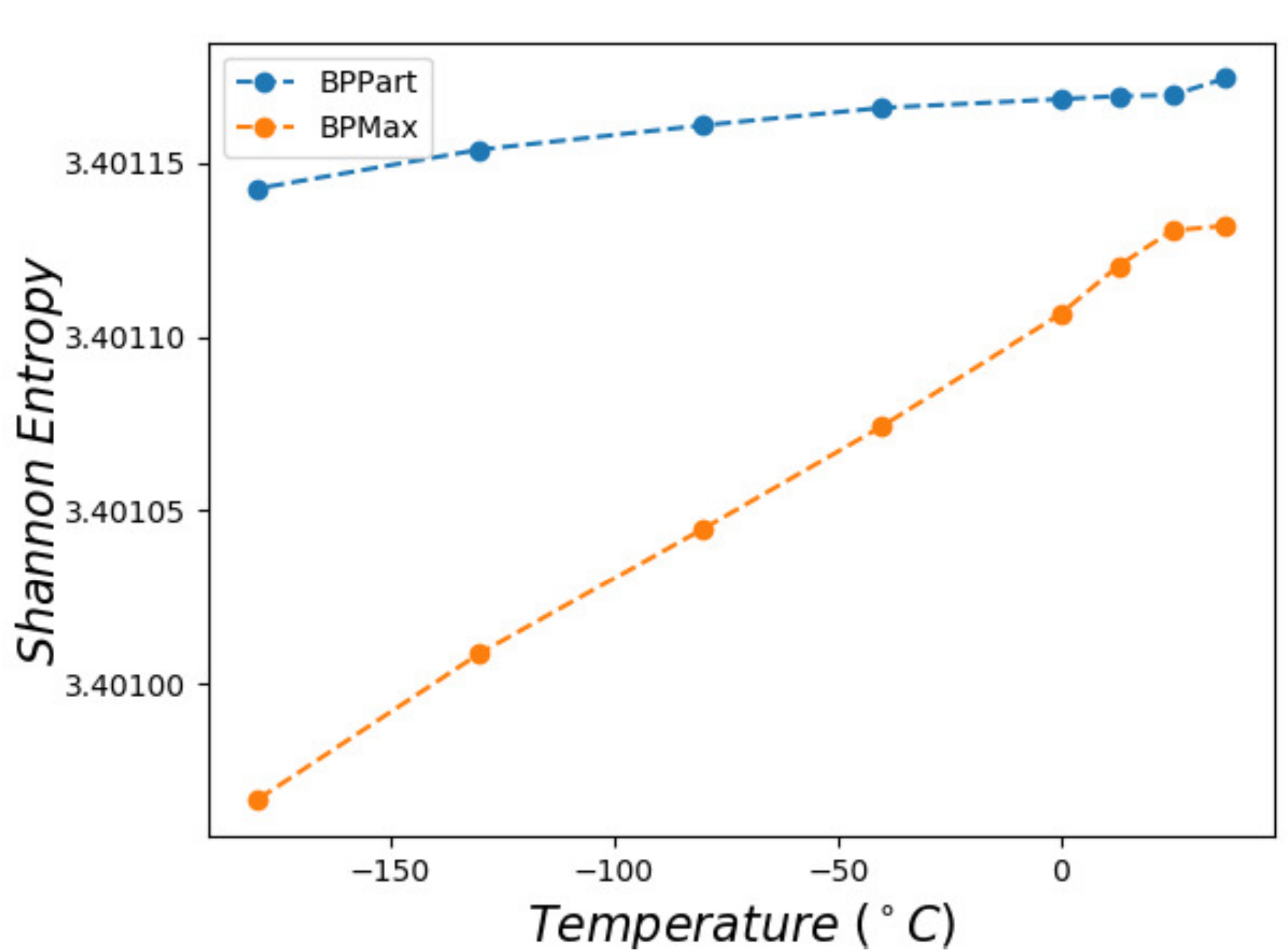}&
\includegraphics[angle=-90,width=0.45\textwidth]{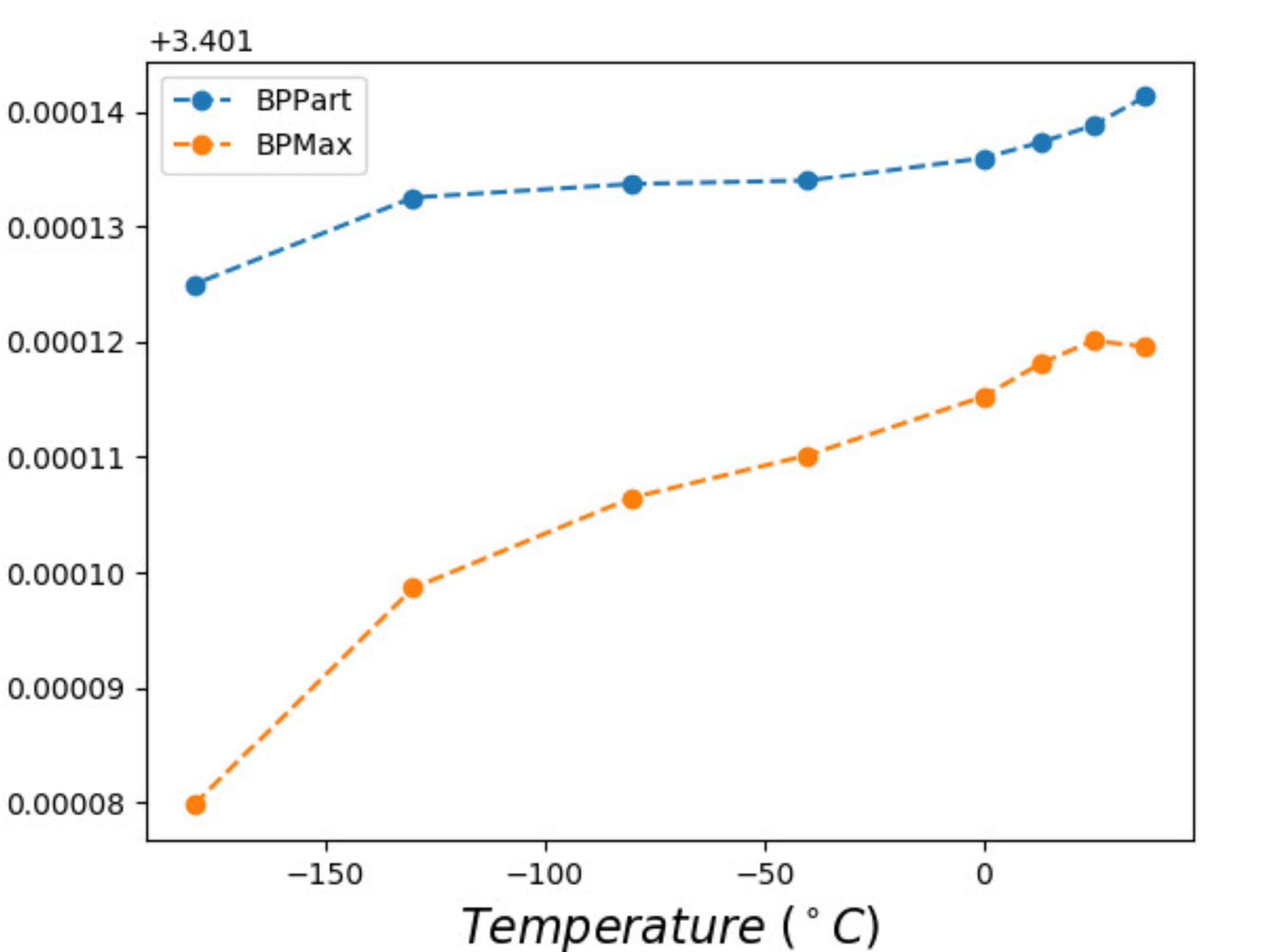}\\
\end{tabular}
\end{center}
\caption{Shannon entropy for the top $30$ Pearson (left) and Spearman's rank
  (right) correlation values at different temperatures for \texttt{BPPart} and
  \texttt{BPMax}.}
\label{fig:ent}
\end{figure*}

Finally to better understand the behavior of the surface around the higher
values in the correlation plots of Figures~\ref{fig:temp-cor_1} and
\ref{fig:temp-cor_2}, we computed the Shannon entropy for the values above a
threshold. Figure~\ref{fig:ent} shows these values for the top $30$ values of
Pearson and Spearman's rank correlation at each temperature.

\section*{Discussion}

The Gibbs free energy
\begin{equation}
\Delta G = \Delta H - T \Delta S
\end{equation}
is composed of a term $\Delta H$ called enthalpy that does not depend on
temperature and a term $T \Delta S$ called entropy that linearly depends on
temperature $T$.  Intuitively, enthalpy is the chemical energy that is often
released upon formation of chemical bonds such as base pairing.  Entropy, on
the other hand, captures the size of all possible spatial conformations for a
fixed secondary structure.  In other words, entropy captures the amount of 3D
freedom of the molecule.  A base pair brings enthalpy down, hence favorable
from enthalpy point of view, and decreases freedom (entropy), hence
unfavorable from entropy point of view.  These two opposing objectives are
combined linearly through the temperature coefficient.

In the full thermodynamic model, we consider both terms.  In the base pair
counting, we consider only a simplistic enthalpy term.  Partition function for
the full thermodynamic model is

\begin{equation}
\sum_{s\in \mathcal{S}^I} e^{-\Delta G / RT},
\end{equation}
in which $R$ is the gas constant.  Note that 
\begin{equation}
-\dfrac{\Delta G}{T} = -\dfrac{\Delta H}{T} + \Delta S, 
\end{equation}
and as $T \to 0$, $-\Delta H/T \to \infty$ and the contribution of $\Delta S$
is diminished to $0$ since it is finite.  Hence, at low temperatures, the
effect of entropy becomes negligible, and we expect to see strong correlation
between the base pair counting model and full thermodynamic model.


\begin{figure*}[t]
\begin{center}
\begin{tabular}{cc}
\includegraphics[angle=-90,width=0.45\textwidth]{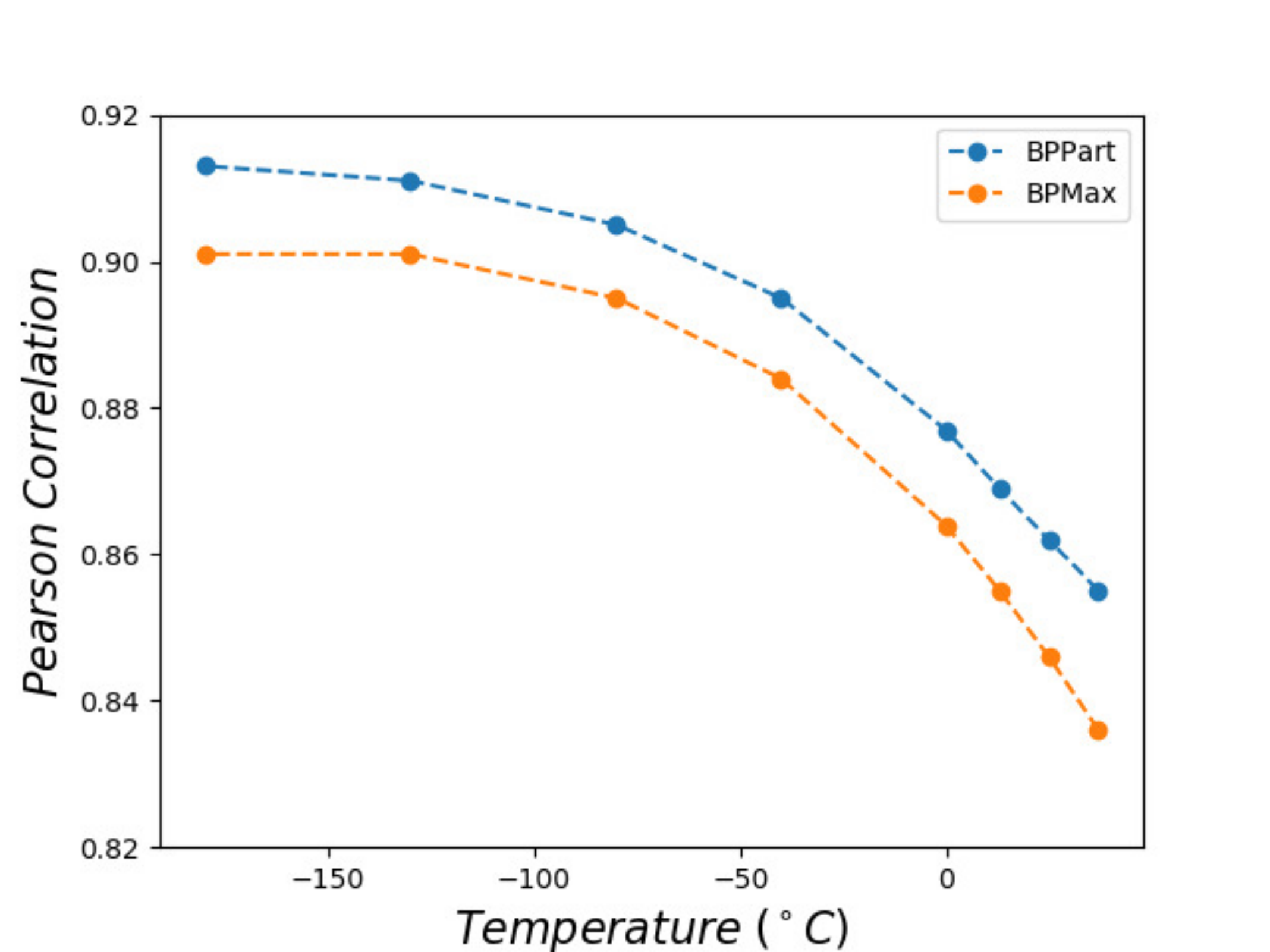}&
\includegraphics[angle=-90,width=0.45\textwidth]{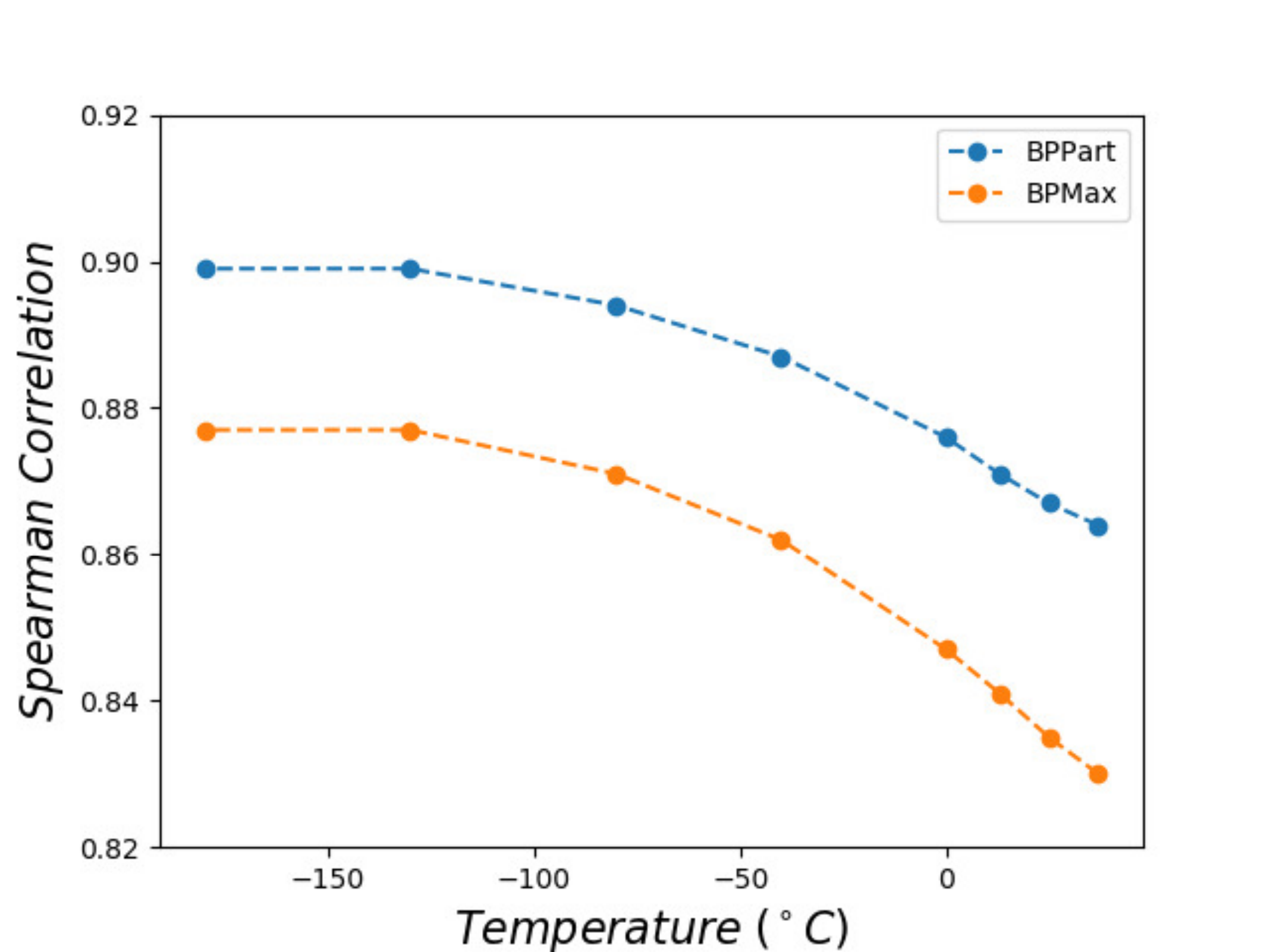}\\
\end{tabular}
\end{center}
\caption{Pearson correlation (left) and Spearman's rank correlation (right)
  between \texttt{piRNA} and \texttt{BPPart} and between \texttt{piRNA} and
  \texttt{BPMax} at different temperatures.}
\label{fig:temp-cor}
\end{figure*}

Figure \ref{fig:temp-cor} shows the Pearson correlations between {\tt BPPart}
 and {\tt BPMax} scores with {\tt piRNA} scores for a fixated combination of
weights that results in the highest correlation at $37\;(^\circ C)$.  For
\texttt{BPPart} the chosen weights are $0.5$, $1.0$, and $3$ for \texttt{AU},
\texttt{GU}, and \texttt{CG}, respectively, while the corresponding weights
for \texttt{BPMax} are $1.0$, $1.5$, and $3$.

Perfectly conforming with the theory, we see higher correlations at low temperatures.  That somewhat validates our implementations as {\tt piRNA} was written totally independently about 10 years ago.  Moreover, as can be seen in Figures \ref{fig:temp-cor_1} and \ref{fig:temp-cor_2}, the surface around the optimum value for higher temperatures becomes flatter.  Figure \ref{fig:ent}, which shows the entropy of the top $30$ correlation values, confirms this observation.  This means the correlation values are less sensitive to a change in the weights of the base pairs as the temperature increases; this conforms with the theory because at higher temperatures, the thermodynamic entropy increases and the total score of \texttt{piRNA} becomes less sensitive to the energy released by pairings.  It is worth mentioning that having less Shannon entropy for the top values at higher temperatures decreases the possibility of having universal optimum values for the weights of the base pairs.  

Another noticeable characteristic of the plots in Figures \ref{fig:temp-cor_1} and
\ref{fig:temp-cor_2} is the region in which the scores of both \texttt{AU} and
\texttt{GU} are non-positive.  This region for \texttt{BPMax} is flat because
when both of these pairs are penalized (or not rewarded when their score is
zero), the algorithm simply avoids making such pairs because it is trying to
maximize the score.  Therefore, it only tries to maximize the number of
\texttt{CG} pairs, which is independent of the scores (penalty in this case) of
the other two types of base pairs.  This also applies to the case where one of
the base pairs has a non-positive score; in that case, \texttt{BPMax} works
independently of the score of that base pair.  So, as soon as any of the
scores becomes non-positive, \texttt{BPMax} remains constant along the
corresponding axis.  For \texttt{BPPart}, however, the story is different
because it simply counts all the possible pairings and even if the score of a
base pair becomes negative, it does not ignore counting that.

Moreover, {\tt BPPart} has a higher correlation than {\tt BPMax} does, which
comes with the price of a $6\times$ increase in empirical running time.  Also, as Figure
\ref{fig:ent} shows, the Shannon entropy for the top $30$ values is less in
\texttt{BPMax} and the gap between them grows as temperature decreases; this
shows that \texttt{BPPart} has a flatter region around the optimum value and
its optimum value is less sensitive to changes in the weights.  Meanwhile,
having a steeper surface in \texttt{BPMax} which has less entropy, increases
the possibility of having more stable and universal optimum values for the
weights.  As mentioned earlier, the running time difference between the two is
noticeable: {\tt BPMax} is about $6\times$ faster than {\tt BPPart}.  Hence,
we now have three choices in increasing order of computational cost:
\texttt{BPMax}, \texttt{BPPart}, and \texttt{piRNA}.  The computation time
increases by about $6\times$~and $225\times$, respectively, from one to the
next.

Finally, based on the results of the experiments on both the primary dataset
and the random one, we see that although the shapes of the optimization plots
and the optimum weights are very similar, the correlation values are much less
for the random dataset.  This observation suggests that probably the
interaction regions are more complementary than the random sequences of the
same size because in these regions the effect of the energy released by
pairing probably becomes more significant than the energy added by an increase
in the entropy on the final score of \texttt{piRNA}.  That could explain why
we get a higher correlation in such regions with \texttt{BPPart} and
\texttt{BPMax}, which mainly depend on the base pairs.  This hypothesis has to
be thoroughly investigated in the future.

\section*{Application of BPPart in Human Biology}

One of the use-cases of \texttt{BPPart} and \texttt{BPMax}, among others, is making predictions about the consequences of a slight change in the RNA sequences. This information becomes helpful for various domains and tasks, such as synthetic biology and studying the mutations. Between \texttt{BPMax} and \texttt{BPPart}, the latter is much more sensitive to small changes in the sequence, because it considers all possible structures that the two interacting sequences might form. Therefore, even a missense mutation might make a tangible difference in the computed \texttt{BPPart} score. 

To verify this hypothesis, we used \texttt{BPPart} to study the effects of known missense mutations, provided by Ensembl, in the interaction regions of some RISE  pairs. Given a pair of interacting RNAs in RISE for which the information about the interacting regions is provided, we retrieved the data of all the reported missense mutations of those regions from Ensembl API. Also, we got the phenotypic consequences of each mutation from Ensembl. Finally, we computed the \texttt{BPPart} score for the original sequence of one of the interacting regions and each of the mutated versions of the other sequence. Among all the generated scores for a pair, we found the outliers using the interquartile range. These outliers, represent a mutation in one of the interacting RNAs, which falls within the interacting region, that causes a great difference in the interaction score. In the rest of this section, we almost-randomly pick and narrate two of such cases that we observed, among many discovered ones. In Appendix A, we report $65$ such pairs that have been discovered using this pipeline after analyzing more than one million pairs of sequences that have been generated after applying the known missense mutations to over $15,200$ pairs of interacting sequences reported in the RISE database. Further study of each of these pairs and more comprehensive study of effect of nonsense mutations on RRI would be a next step in the future.

\subsection*{Traces of TRAF3 in CADASIL}

Cerebral Autosomal Dominant Arteriopathy with Subcortical Infarcts and Leukoencephalopathy (CADASIL) is an inherited condition in which the muscle cells of small blood vessels, especially the ones in the brain, gradually die and cause many impairements, such as stroke, cognitive impairement, and mood disorders in the elderly \cite{di2017cerebral}. It has been shown that mutation in NOTCH3, which resides on the reverse strand of chromosome $19$, is responsible for this condition in people with this genetic disorder \cite{joutel1997notch3}. NOTCH3 and TRAF3 are a pair of interacting RNAs that have been reported in RISE. One of the missense mutations in NOTCH3 that has been reported to be contributing to CADASIL \cite{joutel1996notch3} lies within the interacting region of this gene, from loci $15,161,520$ to $15,161,543$ (according to GRCh38 assembly of human genome), with TRAF3. Interestingly, this mutation, which replaces nucletide $C$ with $G$ at loci $15,161,526$ of chromosome $19$, causes a dramatic increase in the score of \texttt{BPPart} such that it makes it an outlier when the aformentioned procedure is followed. TRAF3 is a gene that has been reported to play a role in angiogenesis \cite{hu1994novel,lalani2015myeloid}. A noticeable increase in the score of \texttt{BPPart} increases the chance that these two RNAs interact and cause post-transcriptional conditions that affect the translation rate of TRAF3 which possibly contributes to the phenotypic consequences of CADASIL. Further evaluation and verification of this hypothesis requires further experimental analysis.

\subsection*{Traces of SNORD3D in Parkinson's Disease}

SNORD3D is a small nucleolar RNA which has been detected not long ago \cite{guttman2009chromatin} with which no specific task or annotation is associated in the literature yet. According to the RISE database, one of the genes that interacts with this snoRNA is GBA which resides on the reverse strand of chromosome $1$. Mutations in GBA has been reported to play a role in Parkinson's disease which is a brain disorder that affects movement and often causes tremors. One of the GBA mutations that is reported to be linked with Parkinson's disease lies within the interaction region of this gene, from loci $155,239,966$ to $155,239,984$ (according to GRCh38 assembly of human genome), with SNORD3D. This specific mutation of GBA, which changes the nucleotide $G$ to $C$ at loci $155,239,972$ of chromosome $1$, is one of the cases that is detected as an outlier using our aforementioned analysis of \texttt{BPPart} scores. This mutation, when applied to GBA, decreases its score of interaction with SNORD3D, which might cause the interaction to occur much less than the normal case. This possibly leads to a change in the expression of GBA protein. According to KEGG, GBA is a member of Other glycan degradation, Sphingolipid metabolism, Metabolic pathways, and Lysosome pathways \cite{Tsuji86}. Therefore, we hypothesize the role of SNORD3D in some or all of those pathways, particularly, the ones that are closely related to Parkinson's disease. Further evaluation of this hypothesis requires further experimental data and analysis.

\section{Conclusion}

We revisited the problems of partition function and structure prediction for
interacting RNAs.  We simplified the energy model and instead considered only
simple weighted base pair counting to obtain \texttt{BPPart} for the partition
function and \texttt{BPMax} for structure prediction.  As a result,
\texttt{BPPart} runs about $225\times$ and \texttt{BPMax} runs about
$1300\times$ faster than \texttt{piRNA} does.  Hence, we gained significant
speedup by potentially sacrificing accuracy.

To evaluate practical accuracy of both new algorithms, we computed the Pearson
and rank correlations at different temperatures between the results of
\texttt{BPPart}, \texttt{BPMax}, and \texttt{piRNA} on 50,500 experimentally
characterized RRIs in the RISE database \cite{Gong17}.  \texttt{BPPart} and
\texttt{BPMax} results correlate well with those of \texttt{piRNA} at low
temperatures.  At the room and body temperatures, there is considerable
correlation and therefore, significant information in the results of
\texttt{BPPart} and \texttt{BPMax}.

We conclude that both \texttt{BPPart} and \texttt{BPMax} capture a significant
portion of the thermodynamic information.  Both tools can be used as filtering
steps in more sophisticated RRI prediction pipelines.  Also, the information
captured by \texttt{BPPart} and \texttt{BPMax} can possibly be complemented
with machine learning techniques in the future for more accurate predictions.
We now have three choices for RRI thermodynamics in increasing computational
cost: {\tt BPMax}, {\tt BPPart}, and {\tt piRNA}.  Depending on the
application and the trade-off between time and accuracy, one can be chosen.

Finally, we show that \texttt{BPPart} might be useful to explain how some mutations lead to some specific phenotypic consequences. We presented two new hypotheses about the roles of TRAF3 in CADASIL and SNORD3D in lipid processing pathways and/or Parkinson's disease.



\subsection{Conflict of interest statement.} None declared.
\newpage

\bibliographystyle{unsrt}
\bibliography{pub,masterref}

\newpage
\eject \pdfpagewidth=17in \pdfpageheight=11.7in

\begin{center}
\noindent {\LARGE \bf Appendix A: 
Disease-causing mutations that incur an outlier \texttt{BPPart} score with a known RRI partner}
\end{center}



\section*{Supplementary material}

disease_causing_outliers

| | rise_id | id1 | id2 | name1 | name2 | mut | chr | strand | position | orig | new | phenotype | IQR | deviation |
|---|---|---|---|---|---|---|---|---|---|---|---|---|---|---|
| 0 | 51273 | ENSG00000136068 | ENSG00000148734 | FLNB | NPFFR1 | 1 | chr3 | + | 58124465 | C | G | FLNB-Related Spectrum Disorders | 1.28964999999999 | -0.0792500000000892 |
| 1 | 50415 | ENSG00000140470 | ENSG00000123607 | ADAMTS17 | TTC21B | 2 | chr2 | - | 165912577 | G | A | Jeune thoracic dystrophy;Joubert Syndrome | 0.894625000000005 | -0.40788749999999396 |
| 2 | 61127 | ENSG00000142197 | ENSG00000105983 | DOPEY2 | LMBR1 | 2 | chr7 | - | 156762153 | A | C | Triphalangeal thumb polysyndactyly syndrome | 1.16395 | 0.257249999999999 |
| 3 | 58037 | ENSG00000136937 | ENSG00000156873 | NCBP1 | PHKG2 | 2 | chr16 | + | 30756397 | T | G | GLYCOGEN STORAGE DISEASE IXc;GLYCOGEN | 0.286300000000004 | 0.872049999999994 |
| 4 | 62929 | ENSG00000189056 | ENSG00000240771 | RELN | ARHGEF25 | 1 | chr7 | - | 103472745 | G | A | Lissencephaly, Recessive | 0.342799999999997 | -2.3163000000000054 |
| 5 | 63228 | ENSG00000117174 | ENSG00000151929 | ZNHIT6 | BAG3 | 2 | chr10 | + | 119672416 | C | G | "Primary familial hypertrophic cardiomyopathy;Myofib | 0.578600000000002 | 0.749600000000001 |
| 6 | 49884 | ENSG00000183943 | ENSG00000156873 | PRKX | PHKG2 | 2 | chr16 | + | 30756397 | T | G | GLYCOGEN STORAGE DISEASE IXc;GLYCOGEN | 0.659300000000002 | 0.744249999999994 |
| 7 | 57967 | ENSG00000158813 | ENSG00000130520 | EDA | LSM4 | 1 | chrX | + | 69957098 | G | T | Hypohidrotic X-linked ectodermal dysplasia;Hyp | 1.0635 | 0.170150000000007 |
| 8 | 54237 | ENSG00000187942 | ENSG00000101463 | LDLRAD2 | SYNDIG1 | 1 | chr1 | + | 21822492 | C | G | Schwartz Jampel syndrome type 1;Dyssegmental Dy | 0.525700000000001 | 2.01465 |
| 9 | 61924 | ENSG00000234465 | ENSG00000156873 | PINLYP | PHKG2 | 2 | chr16 | + | 30756397 | T | G | GLYCOGEN STORAGE DISEASE IXc;GLYCOGEN | 0.400399999999998 | 1.27310000000001 |
| 10 | 53477 | ENSG00000184937 | ENSG00000270641 | WT1 | TSIX | 1 | chr11 | - | 32434971 | A | G | WILMS TUMOR, ANIRIDIA, GENITOURINARY ANO | 0.660525000000007 | 0.515787499999973 |
| 11 | 53142 | ENSG00000177628 | ENSG00000281000 | GBA | SNORD3D | 1 | chr1 | - | 155239972 | G | C | Parkinson disease, late-onset;Parkinson disease, lat | 0.616700000000002 | -0.5511749999999935 |
| 12 | 53382 | ENSG00000128645 | ENSG00000118972 | HOXD1 | FGF23 | 2 | chr12 | - | 4370564 | C | A | HYPOPHOSPHATEMIC RICKETS, AUTOSOMAL D | 0.513374999999996 | -0.1644124999999974 |
| 13 | 58401 | ENSG00000188493 | ENSG00000187535 | C19orf54 | IFT140 | 2 | chr16 | - | 1520311 | G | A | Renal dysplasia, retinal pigmentary dystrophy, cereb | 0.192399999999992 | -1.6374500000000154 |
| 14 | 51135 | ENSG00000161594 | ENSG00000125730 | KLHL10 | C3 | 2 | chr19 | - | 6709754 | G | A | HEMOLYTIC UREMIC SYNDROME, ATYPICAL, SU | 0.292400000000001 | -0.07389999999999475 |
| 15 | 54836 | ENSG00000196712 | ENSG00000163626 | NF1 | COX18 | 1 | chr17 | + | 31225205 | G | A | Neurofibromatosis, type 1 | 0.667774999999999 | -0.31838750000000715 |
| 16 | 51249 | ENSG00000163513 | ENSG00000171444 | TGFBR2 | MCC | 1 | chr3 | + | 30606953 | G | T | Thoracic aortic aneurysm and aortic dissection | 1.158175 | -0.15768750000000153 |
| 17 | 50639 | ENSG00000147027 | ENSG00000164190 | TMEM47 | NIPBL | 2 | chr5 | + | 36876831 | G | C | Cornelia de Lange syndrome | 1.12332499999999 | -1.533412500000043 |
| 18 | 65472 | ENSG00000240490 | ENSG00000177098 | RN7SL277P | SCN4B | 2 | chr11 | - | 118135693 | G | A | Long QT syndrome;Romano-Ward syndrome | 0.259449999999994 | -1.418025000000072 |
| 19 | 52715 | ENSG00000177663 | ENSG00000196943 | IL17RA | NOP9 | 1 | chr22 | + | 17102166 | G | A | Familial Candidiasis, Recessive | 0.998850000000004 | -0.09147499999998843 |
| 20 | 53657 | ENSG00000186684 | ENSG00000156873 | CYP27C1 | PHKG2 | 2 | chr16 | + | 30756397 | T | G | GLYCOGEN STORAGE DISEASE IXc;GLYCOGEN | 0.411999999999999 | 1.0973 |
| 21 | 62243 | ENSG00000197430 | ENSG00000179915 | OPALIN | NRXN1 | 2 | chr2 | - | 51028694 | A | C | Pitt-Hopkins-like syndrome | 0.123999999999995 | -0.5340000000000131 |
| 22 | 55306 | ENSG00000174469 | ENSG00000112319 | CNTNAP2 | EYA4 | 1 | chr7 | + | 148415896 | C | A | Pitt-Hopkins-like syndrome;CORTICAL DYSPLAS | 0.457799999999999 | -0.6374000000000066 |
| 23 | 60574 | ENSG00000166813 | ENSG00000147144 | KIF7 | CCDC120 | 1 | chr15 | - | 89642233 | G | C | Acrocallosal syndrome, Schinzel type" | 1.11322500000001 | 0.218512499999989 |
| 24 | 54812 | ENSG00000054654 | ENSG00000239900 | SYNE2 | ADSL | 2 | chr22 | + | 40364966 | G | A | ADENYLOSUCCINASE DEFICIENCY;Adenylosucci | 1.1359 | -0.3874499999999941 |
| 25 | 49578 | ENSG00000138095 | ENSG00000156521 | LRPPRC | TYSND1 | 1 | chr2 | - | 43887016 | G | A | Leigh syndrome | 0.099999999999994 | -1.6709000000000103 |
| 26 | 50039 | ENSG00000074181 | ENSG00000131323 | NOTCH3 | TRAF3 | 1 | chr19 | - | 15161526 | C | G | Cerebral autosomal dominant arteriopathy with subc | 0.729074999999995 | 0.891187500000015 |
| 27 | 57248 | ENSG00000151929 | ENSG00000279659 | BAG3 | RP11-177G23 | 1 | chr10 | + | 119672297 | C | G | Myofibrillar myopathy, BAG3-related;Dilated Cardio | 0.253600000000006 | 1.49602499999999 |
| 28 | 50475 | ENSG00000141576 | ENSG00000166147 | RNF157 | FBN1 | 2 | chr15 | - | 48432910 | G | A | MARFAN SYNDROME | 0.906750000000002 | -0.240524999999991 |
| 29 | 50754 | ENSG00000124155 | ENSG00000164588 | PIGT | HCN1 | 2 | chr5 | - | 45262033 | G | C | Early infantile epileptic encephalopathy | 0.498525000000001 | -0.038812500000006 |
| 30 | 62856 | ENSG00000130234 | ENSG00000169604 | ACE2 | ANTXR1 | 2 | chr2 | + | 69152194 | G | A | HEMANGIOMA, CAPILLARY INFANTILE, SUSCEPT | 0.279900000000005 | -1.35504999999999 |
| 31 | 61544 | ENSG00000221869 | ENSG00000166813 | CEBPD | KIF7 | 2 | chr15 | - | 89648308 | A | G | Acrocallosal syndrome, Schinzel type" | 0.550025000000005 | 0.614062499999989 |
| 32 | 54646 | ENSG00000127914 | ENSG00000229807 | AKAP9 | XIST | 1 | chr7 | + | 92002229 | G | A | Long QT syndrome | 0.934375000000003 | -0.019387499999994 |
| 33 | 48374 | ENSG00000100345 | ENSG00000178996 | MYH9 | SNX18 | 1 | chr22 | - | 36387950 | C | G | MYH9-related disorder;Nonsyndromic Hearing Loss, | 0.480525 | 0.739812499999999 |
| 34 | 57452 | ENSG00000101974 | ENSG00000156709 | ATP11C | AIFM1 | 2 | chrX | - | 130131756 | G | A | Deafness, X-linked 5" | 0.576374999999999 | -0.539487500000007 |
| 35 | 64673 | ENSG00000075624 | ENSG00000180182 | ACTB | MED14 | 1 | chr7 | - | 5529594 | G | A | BARAITSER-WINTER SYNDROME 1 | 0.971800000000002 | -0.159500000000008 |
| 36 | 64673 | ENSG00000075624 | ENSG00000180182 | ACTB | MED14 | 1 | chr7 | - | 5529624 | A | G | BARAITSER-WINTER SYNDROME 1;BARAITSER- | 0.971800000000002 | 0.399899999999988 |
| 37 | 64673 | ENSG00000075624 | ENSG00000180182 | ACTB | MED14 | 1 | chr7 | - | 5529624 | A | C | BARAITSER-WINTER SYNDROME 1;BARAITSER- | 0.971800000000002 | 0.14139999999999 |
| 38 | 51417 | ENSG00000164619 | ENSG00000105576 | BMPER | TNPO2 | 1 | chr7 | + | 33905094 | G | A | Diaphanospondylodysostosis | 0.296549999999996 | -0.386200000000017 |
| 39 | 62328 | ENSG00000232316 | ENSG00000170289 | RP1-124C6 | CNGB3 | 2 | chr8 | - | 86574166 | G | A | Stargardt Disease, Recessive;Achromatopsia | 0.72547500000001 | -0.24951249999998 |
| 40 | 53469 | ENSG00000183196 | ENSG00000166233 | CHST6 | ARIH1 | 1 | chr16 | - | 75477044 | T | G | Macular corneal dystrophy Type I | 0.450999999999993 | 0.446200000000005 |
| 41 | 61935 | ENSG00000183873 | ENSG00000108306 | SCN5A | FBXL20 | 1 | chr3 | - | 38550198 | A | G | Paroxysmal familial ventricular fibrillation;Roman | 0.237174999999993 | 1.6237375 |
| 42 | 59210 | ENSG00000207110 | ENSG00000173575 | RNU1-106P | CHD2 | 2 | chr15 | + | 93020077 | G | A | Epileptic encephalopathy, childhood-onset;Epileptic | 0.112300000000005 | -0.015799999999992 |
| 43 | 64224 | ENSG00000165283 | ENSG00000183873 | STOML2 | SCN5A | 2 | chr3 | - | 38633255 | G | A | Congenital long QT syndrome | 0.395350000000008 | -0.459074999999984 |
| 44 | 48691 | ENSG00000110092 | ENSG00000076248 | CCND1 | UNG | 2 | chr12 | + | 109110470 | C | G | Immunodeficiency with Hyper-IgM | 0.374000000000002 | 1.55159999999999 |
| 45 | 62535 | ENSG00000105576 | ENSG00000110799 | TNPO2 | VWF | 2 | chr12 | - | 5981833 | G | T | von Willebrand disorder | 0.867199999999997 | -0.523500000000013 |
| 46 | 62535 | ENSG00000105576 | ENSG00000110799 | TNPO2 | VWF | 2 | chr12 | - | 5981833 | G | A | von Willebrand disorder | 0.867199999999997 | -1.13360000000001 |
| 47 | 51170 | ENSG00000149136 | ENSG00000100234 | SSRP1 | TIMP3 | 2 | chr22 | + | 32862388 | G | A | Pseudoinflammatory fundus dystrophy | 0.221199999999996 | -1.61210000000001 |
| 48 | 58057 | ENSG00000185920 | ENSG00000162341 | PTCH1 | TPCN2 | 1 | chr9 | - | 95469067 | A | G | Gorlin syndrome | 0.730800000000016 | 1.06344999999997 |
| 49 | 58057 | ENSG00000185920 | ENSG00000162341 | PTCH1 | TPCN2 | 1 | chr9 | - | 95469088 | G | A | Gorlin syndrome | 0.730800000000016 | -0.740349999999964 |
| 50 | 53161 | ENSG00000183864 | ENSG00000075624 | TOB2 | ACTB | 2 | chr7 | - | 5529624 | A | C | BARAITSER-WINTER SYNDROME 1;BARAITSER- | 0.956874999999997 | 0.639812500000005 |
| 51 | 64674 | ENSG00000075624 | ENSG00000251497 | ACTB | RP11-197N18 | 1 | chr7 | - | 5529594 | G | A | BARAITSER-WINTER SYNDROME 1 | 0.705799999999996 | -0.336849999999998 |

disease_causing_outliers

| | | | | | | | | | | | | |
|---|---|---|---|---|---|---|---|---|---|---|---|---|
| 52 | 53476 | ENSG00000184937 | ENSG00000252680 | WT1 | RNA5SP449 | 1 | chr11 | - | 32434971 | A | G | WILMS TUMOR, ANIRIDIA, GENITOURINARY ANO | 0.821150000000003 | 0.399924999999982 |
| 53 | 53476 | ENSG00000184937 | ENSG00000252680 | WT1 | RNA5SP449 | 1 | chr11 | - | 32434980 | C | G | Drash syndrome;WILMS TUMOR, ANIRIDIA, GEN | 0.821150000000003 | 0.681824999999989 |
| 54 | 65331 | ENSG00000163686 | ENSG00000075624 | ABHD6 | ACTB | 2 | chr7 | - | 5527786 | G | A | BARAITSER-WINTER SYNDROME 1 | 0.394350000000003 | -0.161524999999998 |
| 55 | 51773 | ENSG00000164588 | ENSG00000105576 | HCN1 | TNPO2 | 1 | chr5 | - | 45695893 | T | C | Early infantile epileptic encephalopathy | 0.37462499999998 | 0.822012500000042 |
| 56 | 53048 | ENSG00000158467 | ENSG00000105610 | AHCYL2 | KLF1 | 2 | chr19 | - | 12885877 | G | A | Congenital dyserythropoietic anemia | 0.503799999999998 | -0.714799999999997 |
| 57 | 61065 | ENSG00000249158 | ENSG00000130164 | PCDHA11 | LDLR | 2 | chr19 | + | 11132534 | C | G | Familial hypercholesterolemia | 0.647699999999986 | 0.641475000000028 |
| 58 | 53557 | ENSG00000184634 | ENSG00000140323 | MED12 | DISP2 | 1 | chrX | + | 71137822 | G | T | FG syndrome | 0.388400000000004 | -1.47449999999998 |
| 59 | 58063 | ENSG00000072501 | ENSG00000202324 | SMC1A | RNA5SP366 | 1 | chrX | - | 53376265 | G | A | Cornelia de Lange syndrome | 0.564299999999989 | -0.531150000000011 |
| 60 | 56554 | ENSG00000201861 | ENSG00000136068 | RNA5SP298 | FLNB | 2 | chr3 | + | 58077256 | G | A | FLNB-Related Disorders | 0.698800000000006 | -1.1193 |
| 61 | 56554 | ENSG00000201861 | ENSG00000136068 | RNA5SP298 | FLNB | 2 | chr3 | + | 58077266 | T | G | BOOMERANG DYSPLASIA;BOOMERANG DYSPLA | 0.698800000000006 | 0.63339999999998 |
| 62 | 56554 | ENSG00000201861 | ENSG00000136068 | RNA5SP298 | FLNB | 2 | chr3 | + | 58077271 | G | A | ATELOSTEOGENESIS, TYPE III;Atelosteogenesis t | 0.698800000000006 | -0.962399999999988 |
| 63 | 53823 | ENSG00000124107 | ENSG00000169071 | SLPI | ROR2 | 2 | chr9 | - | 91733195 | G | A | Brachydactyly;Robinow syndrome | 0.744950000000003 | -0.730199999999982 |
| 64 | 49653 | ENSG00000101577 | ENSG00000198931 | LPIN2 | APRT | 1 | chr18 | - | 2920313 | G | A | MAJEED SYNDROME | 0.404200000000003 | -1.661575 |



\end{document}